\documentclass{aa}  
\usepackage{graphicx}
\usepackage{longtable}
\usepackage{lscape}
\usepackage{subfig}
\usepackage{float}
\usepackage{natbib}
\usepackage{multirow}
\usepackage{graphicx}
\usepackage{txfonts}
\begin{document}

   \title{Two-dimensional multi-component photometric decomposition of CALIFA
     galaxies}


   \author{J. M\'endez-Abreu \inst{1}
          \and
T. Ruiz-Lara \inst{2,3}
\and
L. S\'anchez-Menguiano \inst{2,4}
\and
A. de Lorenzo-C\'aceres \inst{1,2}
\and
L. Costantin \inst{5}
\and
C. Catal\'an-Torrecilla \inst{6}
\and
E. Florido \inst{2,3}
\and
J. A. L. Aguerri \inst{7,8}
\and
J. Bland-Hawthorn \inst{9} 
\and
E. M. Corsini \inst{5,10}
\and
R. J. Dettmar \inst{11}
\and
L. Galbany \inst{12,13}
\and
R. Garc\'ia-Benito \inst{4}
\and
R. A. Marino \inst{14} 
\and
I. M\'arquez \inst{4}
\and 
R. A. Ortega-Minakata \inst{15}
\and
P. Papaderos \inst{16}
\and
S. F. S\'anchez \inst{17}
\and
P. S\'anchez-Blazquez \inst{18,19}
\and
K. Spekkens \inst{20}
\and
G. van de Ven \inst{21}
\and
V. Wild \inst{1}
\and
B. Ziegler \inst{22}
          }

\institute{School of Physics and Astronomy, University of St. Andrews, SUPA, North Haugh, KY16 9SS, St. Andrews, UK\\
              \email{jma20@st-andrews.ac.uk}
\and
Departamento de F\'isica Te\'orica y del Cosmos, Universidad de Granada, Campus de Fuentenueva, E-18071 Granada, Spain
\and
Instituto Carlos I de F\'isica Te\'orica y Computacional, Universidad de Granada, E-18071 Granada, Spain
\and
Instituto de Astrof\'isica de Andaluc\'ia (CSIC), Glorieta de la Astronom\'ia s/n, E-3004, E-18080 Granada, Spain
\and
Dipartimento di Fisica e Astronomia ‘G. Galilei’, Universit\`a di Padova, vicolo dell’Osservatorio 3, I-35122 Padova, Italy
\and
Departamento de Astrof\'isica y CC. de la Atm\'osfera, Universidad Complutense de Madrid, E-28040 Madrid, Spain
\and 
Instituto de Astrof\'isica de Canarias, Calle V\'ia L\'actea s/n, E-38205 La Laguna, Tenerife, Spain
\and
Departamento de Astrof\'isica, Universidad de La Laguna, E-38200 La Laguna, Tenerife, Spain
\and
Sydney Institute for Astronomy, School of Physics A28, University of Sydney, NSW 2006, Australia
\and
INAF–Osservatorio Astronomico di Padova, vicolo dell’Osservatorio 5, I-35122 Padova, Italy
\and
Astronomisches Institut, Ruhr-Universit\"at Bochum, Universit\"atsstr. 150, D-44801 Bochum, Germany
\and
Pittsburgh Particle Physics, Astrophysics, and Cosmology Center (PITT PACC)
\and
Physics and Astronomy Department, University of Pittsburgh, Pittsburgh, PA 15260, USA
\and
ETH Z\"urich, Institute for Astronomy, Wolfgang-Pauli-Str. 27, 8093 Z\"urich, Switzerland
\and
Observat\'orio do Valongo, Universidade Federal do Rio de Janeiro, Ladeira do Pedro Ant\^onio 43, Sa\'ude, Rio de Janeiro, RJ 20080-090, Brazil
\and
Instituto de Astrofísica e Ci\^encias do Espa\c{c}o, Universidade do Porto, Centro de Astrof{\'\i}sica da Universidade do Porto, Rua das Estrelas, 4150-762 Porto, Portugal
\and
Instituto de Astronom\'ia, Universidad Nacional Aut\'onoma de M\'exico, A.P. 70-264, 04510 M\'exico D.F., Mexico
\and
Departamento de F\'isica Te\'orica, Universidad Aut\'onoma de Madrid, E-28049 Cantoblanco, Spain
\and
Instituto de Astrofísica, Pontificia Universidad Católica de Chile, Av. Vicuña Mackenna 4860, 782-0436 Macul, Santiago, Chile
\and
Department of Physics, Royal Military College of Canada, P.O. Box 17000, Station Forces, Kingston, Ontario K7K 7B4, Canada
\and
Max-Planck-Institut f\"ur Astronomie, K\"onigstuhl 17, D-69117 Heidelberg, Germany
\and
University of Vienna, Department of Astrophysics, T\"urkenschanzstrasse 17, 1180 Vienna, Austria
             }

   \date{Received September 15, 1996; accepted March 16, 1997}


 \abstract
{
We present a two-dimensional multi-component photometric decomposition
of 404  galaxies from the Calar  Alto Legacy Integral Field  Area Data
Release 3 (CALIFA-DR3).  They represent  all possible galaxies with no
clear  signs of  interaction and  not strongly  inclined in  the final
CALIFA data release.   Galaxies are modelled in the $g$,  $r$, and $i$
Sloan Digital Sky Survey (SDSS)  images including, when appropriate, a
nuclear point  source, bulge, bar,  and an exponential or  broken disc
component.   We  use  a  human-supervised approach  to  determine  the
optimal number of structures to be  included in the fit.  The dataset,
including the photometric parameters of the CALIFA sample, is released
together with statistical errors and  a visual analysis of the quality
of each fit.
The analysis of the photometric components reveals a clear segregation
of the structural  composition of galaxies with stellar  mass. At high
masses (${\rm log(M_{\star}/M_{\sun})}>11$),  the galaxy population is
dominated by galaxies modelled with  a single S\'ersic or a bulge+disc
with  a   bulge-to-total  ($B/T$)  luminosity  ratio   $B/T>0.2$.   At
intermediate masses ($9.5<{\rm log(M_{\star}/M_{\sun})}<11$), galaxies
described with bulge+disc  but $B/T < 0.2$  are preponderant, whereas,
at  the  low  mass   end  (${\rm  log(M_{\star}/M_{\sun})}<9.5$),  the
prevailing population is constituted  by galaxies modelled with either
pure  discs  or  nuclear  point sources+discs  (i.e.,  no  discernible
bulge).
We obtain that 57\% of the volume corrected sample of disc galaxies in
the CALIFA sample  host a bar.  This bar fraction  shows a significant
drop   with   increasing   galaxy   mass  in   the   range   $9.5<{\rm
  log(M_{\star}/M_{\sun})}<11.5$.
The analyses of the extended multi-component radial profile result in
a  volume-corrected distribution  of  62\%,  28\%,  and 10\%  for  the
so-called Type I (pure exponential),  Type II (down-bending), and Type
III (up-bending)  disc profiles, respectively. These  fractions are in
discordance  with  previous findings.   We  argue  that the  different
methodologies used to  detect the breaks are the main  cause for these
differences.

%
%
%
%
%
%
%
%
%
%
%
}


   \maketitle
\section{Introduction}
\label{sec:intro}

The constant development of  the morphological classification schemes,
from  the Hubble  tuning  fork diagram  \citep{hubble26}  to the  more
sophisticated  Comprehensive  de  Vaucouleurs  revised  Hubble-Sandage
\citep[CVRHS;][]{buta15}   catalogue,  illustrate   the  morphological
complexity  of galaxy  systems.  Even  apparently simple  systems like
ellipticals can host a wealth of other structures such as outer shells
or  nuclear cores  \citep{malin80,morelli04}.  The  properties of  the
different  stellar structures  that  make up  galaxies (e.g.,  bulges,
bars,  and  discs)  are  the  direct result  of  their  formation  and
evolution.   Therefore,  the  quantification   of  the  properties  of
galaxies and their  distinct stellar structures is  a fundamental step
towards understanding how galaxies form and evolve.

Historically, since the  early era of photographic plates,  one of the
key methods for studying the  projected luminosity density, or surface
brightness (SB), of galaxies was to use parametric laws to model their
different components \citep{freeman70,kormendy77}.   In the early 80s,
the  pioneering  photometric  decomposition   methods  were  based  on
modelling the one-dimensional (1D) SB  profiles of galaxies as the sum
of             separate             components             \citep[e.g.
  bulge+disc;][]{boroson81,send82,hickson82}.   With   the  advent  of
charge couple device (CCD) cameras, these first attempts to understand
the stellar structures  that shape a galaxy were improved  upon in the
following decades by  hybrid methods fitting not only the  SB but also
the  galaxy ellipticity  profiles \citep{trujillo01b,aguerri02}.   The
first  pixel-by-pixel fitting  of galaxy  images with  two-dimensional
(2D) SB models was done  by \citet{byunfreeman95}. Since then, several
codes  have  been  developed  to  perform  2D  bulge+disc  photometric
decompositions  such  as  GIM2D   \citep{simard98},  but  most  recent
algorithms allow  for a  2D multi-component  photometric decomposition
(e.g.  GALFIT, \citealt{peng02},  BUDDA, \citealt{desouza04},  GASP2D,
\citealt{mendezabreu08}, IMFIT, \citealt{erwin15}).

The  necessity  for  a  multi-component approach  to  the  photometric
decomposition  of  galaxies has  been  demonstrated  in several  works
\citep{prieto01,gadotti08,  salo15}. In  particular, the  inclusion of
the bar component  has been proved to be critical  in order to recover
accurate  bulge  parameters  \citep[e.g.][]{aguerri05,laurikainen05}.
Several studies  have shown that both  the Sérsic index ($n$)  and the
bulge-to-total luminosity ratio ($B/T$)  can be artificially increased
if   the   bar   is   not   properly  accounted   for   in   the   fit
\citep{gadotti09,weinzirl09,mendezabreu14}.    In   addition,   recent
advances  in  observational  techniques   have  allowed  for  detailed
analysis  of  the  light  distribution in  the  outermost  regions  of
galaxies.   Despite  the  classical  view  of  galaxies  hosting  pure
exponential  discs  being  confirmed   with  ultra  deep  observations
\citep[e.g.][]{blandhawthorn05},  a wide  variety  of outer  profiles
deviating   from  a   pure   exponential  have   also  been   reported
\citep{erwin05,  pohlentrujillo06}.  Therefore,  since  the origin  of
breaks in disc  profiles is still debated,  a multi-component approach
taking  into  account  broken  profiles is  of  vital  importance  for
understanding disc formation  and evolution processes \citep{marino16,
  ruizlara16}.

Despite  their  limitations,  two-component  (bulge+disc)  photometric
decompositions are still the common  procedure when dealing with large
surveys         at        low         and        high         redshift
\citep{allen06,simard11,lacknergunn12,haussler13}.   This   is  mainly
because current methodologies to find the best fit model to the galaxy
images   using   two-component   models   are   relatively   easy   to
automatise. However,  when more  structures are  added to  the fitting
process, they become more degenerate  and human supervision is usually
needed.  A number of studies have attempted to produce multi-component
photometric  decompositions  of  samples   with  several  hundreds  of
galaxies.     Recently,   \citet{salo15}    performed   the    largest
multi-component decomposition  to date,  analysing 2352  galaxies from
the    Spitzer   Survey    of    Stellar    Structure   in    Galaxies
\citep[S$^4$G,][]{sheth10}.

Here, we  present the 2D multi-component  photometric decomposition of
404  galaxies drawn  from the  final data  release of  the Calar  Alto
Legacy Integral  Field Area survey  \citep[CALIFA,][]{sanchez12}.  The
galaxy sample  represents all galaxies  in the CALIFA survey  that are
suitable  for our  photometric  analysis,  that is,  they  are not  in
interaction  with other  galaxies and  are not  heavily inclined.  The
CALIFA survey  is an integral  field spectroscopy (IFS) survey  of 667
galaxies that provides spatially  resolved information such as stellar
and  gas  kinematics,  stellar  populations,  and  gas-phase  physical
properties  over  a   large  field  of  view   (2-3  galaxy  effective
radii). The CALIFA data  have significantly improved our understanding
of  the  physical processes  leading  to  the observed  population  of
galaxies in the  nearby Universe. The aim of this  paper is to provide
the CALIFA  dataset with  an accurate photometric  characterisation of
the multiple  stellar structures shaping the  CALIFA galaxies (bulges,
bars, and  discs). To this aim,  we have used the  homogeneous imaging
provided   by  the   Sloan   Digital  Sky   Survey   Data  Release   7
\citep[SDSS-DR7]{abazajian09} for  the whole  CALIFA sample.   Some of
the information presented here has already been used within the survey
collaboration to  study the  influence of  bars in  stellar population
gradients   \citep{sanchezblazquez14},   to    address   the   stellar
populations   of  different   disc  profiles   \citep{ruizlara16},  to
understand       gas      abundance       gradient      in       discs
\citep{sanchez14,sanchezmenguiano16},     and    to     analyse    the
morpho-kinematic   properties  of   bulges   in  lenticular   galaxies
(Mendez-Abreu et al.  submitted). This  paper focuses on the technical
aspects  of the  photometric decomposition  and the  incidence of  the
different galaxy structures in the  CALIFA galaxy sample. The detailed
photometric  description of  the galaxy  structures presented  in this
paper opens a  new set of possibilities to the  wealth of 2D spatially
resolved spectroscopic information provided  by the CALIFA survey. The
properties of the individual structures  and their relation with other
galaxy properties extracted  from the CALIFA data will  be explored in
future papers.

The  current paper is  organised  as follows:  Sect.~\ref{sec:sampleselection}
describes the CALIFA data release 3 (DR3) and the final sample used in
this study.  Sect.~\ref{sec:photdec} details  the technical aspects of
the 2D photometric  decomposition analysis.  Sect.~\ref{sec:multiphot}
describes the  fitting process and  the main types  of multi-component
decomposition carried  out in  this paper. The  incidence of  the main
stellar structures found in our sample  are analysed in the context of
the global properties  of the galaxies.  Sect.~\ref{sec:uncertainties}
presents a complete  description of the uncertainties  inherent to our
analysis.  The  conclusions are given  in Sect.~\ref{sec:conclusions}.
Throughout the paper we assume  a flat cosmology with $\Omega_{\rm m}$
= 0.3, $\Omega_{\rm \Lambda}$ = 0.7,  and a Hubble constant $H_0$ = 70
km s$^{-1}$ Mpc$^{-1}$.

\section{The CALIFA DR3 and our sample selection}
\label{sec:sampleselection}

\begin{figure*}[!ht]
\includegraphics[bb=54 345 558 500]{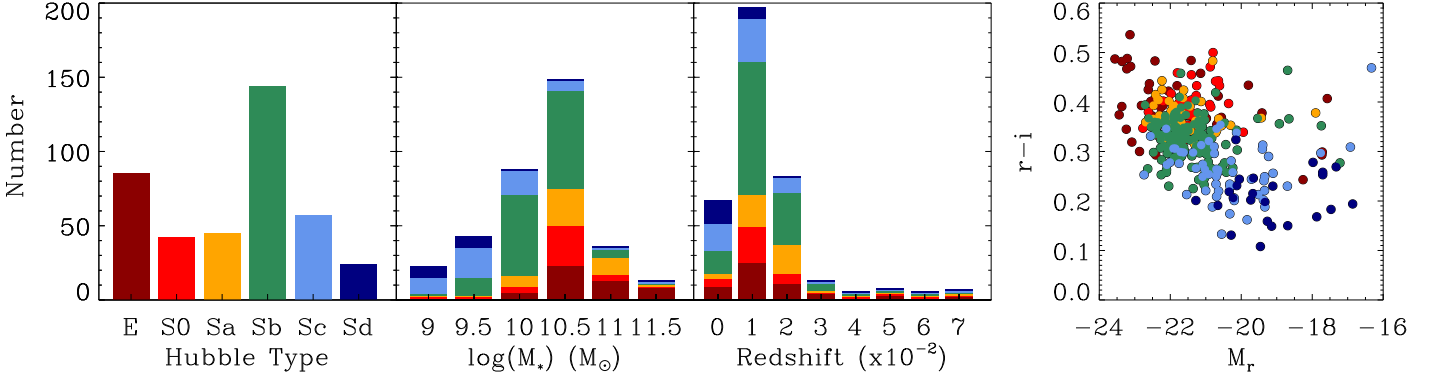}
\caption{({\it from left to right})  Distribution of our galaxy sample
  in  Hubble  type, stellar  mass,  redshift,  and  on the  $r-i$  vs.
  $r-$band colour  magnitude relation.  Colours in  all panels represent
  the different galaxy  Hubble types.  The parameters  for each galaxy
  were obtained from \citet{walcher14}.}
\label{fig:sample}
\end{figure*}

The CALIFA final data  release \citep[DR3,][]{sanchez16} comprises two
different samples of galaxies: galaxies belonging to the CALIFA mother
sample and  galaxies that  are considered  the extension  sample.  The
first group represents the natural expansion of the galaxies presented
in   the    previous   CALIFA    DR1   \citep{husemann13}    and   DR2
\citep{garciabenito15}, and fully  characterised in \citet{walcher14}.
The  second group  corresponds to  a compendium  of different  sets of
galaxies that were observed using the same setup as CALIFA, as part of
ancillary science projects within  the CALIFA collaboration. The final
CALIFA DR3 comprises 667 galaxies.

From the final data release,  and after performing a visual inspection
of the SDSS images, we carry  out a discard/exclusion process of those
galaxies not  suitable for  our photometric  study.  First,  we reject
paired and interacting objects (57  galaxies) as well as those systems
with a  heavily distorted morphology  (19 galaxies).  Since we  aim to
provide an accurate description of the galaxy stellar structures using
only  symmetric models,  galaxies  with distorted  features cannot  be
successfully  modelled.  Then,  we check  for the  presence of  bright
stars  contaminating our  target  galaxies and  remove  them from  the
analysis  (5 galaxies).   Finally, the  identification and  subsequent
characterisation of structures in  highly inclined galaxies is usually
not possible due  to projection effects, thus those  galaxies close to
edge-on  ($i >  70^{\circ}$) were  also removed  from the  sample (183
galaxies).  This latter  process was performed by  a visual inspection
of individual  galaxies since we find  that a typical disc  axis ratio
cut  does not  work  for early-type  edge-on  galaxies with  spherical
stellar haloes.  The final sample presented in this paper contains 404
galaxies.   The  distribution  of their  main  global  characteristics
extracted from \citet{walcher14} is shown in Fig.~\ref{fig:sample}.

\begin{figure}[!ht]
\includegraphics[width=0.49\textwidth]{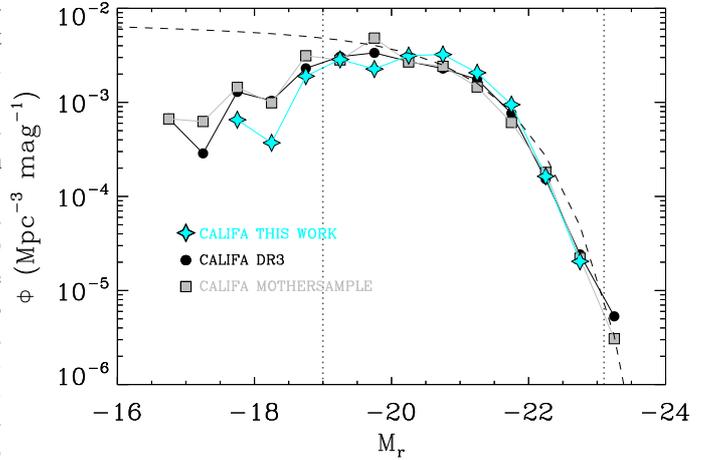}
\caption{Luminosity  functions  of  the  CALIFA  mother  sample  (grey
  squares), final CALIFA observed sample  in the DR3 (black dots), and
  the sample  described in this  paper (cyan stars). The  dotted lines
  denote the  mother sample completeness limits.   The SDSS luminosity
  function of \citet{blanton03} is shown with a black dashed line.}
\label{fig:lf}
\end{figure}

The CALIFA  mother sample presents the  noticeable characteristic that
its selection  criteria are well understood.   Therefore, although the
final observed  sample is not  complete in  volume, it is  possible to
reconstruct volume corrected sample properties. The complete procedure
is described in  \citet{walcher14} and assigns a  volume correction to
each individual galaxy  that can be used to correct  for the selection
function.   Fig.~\ref{fig:lf} shows  the  luminosity  function of  the
sample in this study, the CALIFA mother sample, and the final observed
sample  in  the  CALIFA  DR3.    It  is  worth  noticing  that  volume
corrections are  not applicable  to the extended  sample due  to their
complicated   sample  selection.    Thus,   the  luminosity   function
represented in  Fig.~\ref{fig:lf} contains  only those  galaxies drawn
from the  CALIFA mother sample (297  galaxies).  We find a  good match
among the luminosity functions of the three different samples, as well
as for the SDSS luminosity function given by \citet{blanton03}, within
the completeness  limits described in \citet{walcher14},  that is, -19
$> M_r>$-23.1. Our sample contains 285 galaxies within these limits.

\section{2D Photometric decomposition analysis: model functions and data reduction}
\label{sec:photdec}

The  structural  parameters  of  the CALIFA  sample  were  derived  by
applying a 2D photometric decomposition. The use of algorithms fitting
the 2D  surface-brightness distributions (SBD) of  galaxies has become
standard practice in the analysis of galaxy photometric structures due
to      the      advantages      over     simple      1D      analysis
\citep[see,][]{byunfreeman95,erwin15}.   In this  paper  we apply  the
GASP2D  code \citep{mendezabreu08,  mendezabreu14}.   GASP2D adopts  a
Leverberg-Marquard  algorithm to  fit the  2D SBD  of galaxies  with a
versatile  set  of  model   components.   This  section  presents  the
analytical functions chosen  to describe each galaxy model  as well as
our pre-processing  of the SDSS images  to be used in  the photometric
decomposition.

\subsection{Analytical functions for the structural components}
\label{sec:function}

The  galaxy SBD  is  assumed to  be the  sum  of multiple  photometric
structures depending  on its  specific morphology.  GASP2D  allows the
simultaneous  fitting of  different galaxy  structures: nuclear  point
source (NPS), bulge, bar, and  disc (including breaks).  Each of these
components is built such that  its corresponding isophotes are perfect
ellipses centred  on $(x_0,\,y_0)$  with constant position  angle (PA)
and  constant ellipticity  ($\epsilon  = 1  - q$),  where  $q$ is  the
minor-to-major axis  ratio of  the ellipse.  The  geometric parameters
(PA, $\epsilon$)  are independent  for each component.   The isophotal
radius $r$ is then given by
 
\begin{eqnarray} 
r & = &\left[(-(x-x_0)\sin{{\rm PA}} + (y-y_0)  
\cos{{\rm PA}})^2 + \right. \nonumber \\  
    &   &  \left.(-(x-x_0)\cos{{\rm PA}} - (y-y_0)  
\sin{{\rm PA}})^2/q^2\right]^{1/2}.  
\label{eqn:bulge_radius} 
\end{eqnarray} 

The  SBD of  the  bulge  component is  parameterised  with a  S\'ersic
profile  \citep{sersic68},   also  known  as  the   $r^{1/n}$  law  or
generalised de Vaucouleurs law.

\begin{equation} 
I_{\rm b}(r_{\rm b})=I_{\rm e}10^{-b_n\left[\left(\frac{r_{\rm b}}{r_{\rm e}} 
\right)^{\frac{1}{n}}-1\right]}, 
\label{eqn:bulge_surfbright} 
\end{equation} 
%
where $r_{\rm b}$ is the  radius measured in the Cartesian coordinates
describing the reference system of the  bulge in the plane of the sky.
$r_{\rm e}$,  $I_{\rm e}$, and  $n$ are the effective  (or half-light)
radius, the surface brightness at  $r_{\rm e}$, and the S\'ersic index
describing the  curvature of  the SBD,  respectively, and  $b_n \simeq
0.868\,n-0.142$ \citep{caon93}.

The NPS is modelled using a Moffat function mimicking the point spread
function  (PSF) of  the galaxy  image. Several  works have  proven the
importance of including a NPS  to properly derive the bulge parameters
when nuclear  stellar clusters  \citep[NSC, ][]{balcells07}  or active
galactic nuclei \citep[AGN, ][]{benitez13} are present. In this paper,
the inclusion  of a NPS  does not  intend to model  an extra component
(NSC,  AGN) but  rather  to  model an  unresolved  bulge  with a  size
comparable  to the  image PSF.   Therefore, the  use of  a NPS  in the
modelling of a galaxy rules out  the simultaneous fitting with a bulge
component  (see  Sect.~\ref{sec:NPS}  for further  details).  The  NPS
parameterisation is given by

\begin{equation} 
I_{\rm NPS}(r_{\rm NPS})=I_{\rm NPS}\,\left(1 + \left(\frac{r_{\rm NPS}}{\alpha}\right)^2\right)^{-\beta}  ,
\label{eqn:psf} 
\end{equation} 
%
where the parameters $\alpha$ and $\beta$ define the profile shape and
are  related  to  the  full  width at  half  maximum  (FHWM)  such  as
FWHM=2$\alpha$$\sqrt{2^{1/\beta}-1}$.

The SBD  of a  galaxy disc  is usually  described with  an exponential
profile. However, nowadays  it is commonly accepted  that galaxy discs
can     be     classified     into    three     general     categories
\citep{erwin05,pohlentrujillo06}: (i)  Type I  profiles that  follow a
single  exponential profile  along  the whole  optical  extent of  the
galaxies, (ii) Type II profiles  that present a double exponential law
with a down-bending beyond the  so-called break radius, and (iii) Type
III profiles  that exhibit  an up-bending  in the  outer parts  of the
discs.  To account  for  these possibilities  we  adopt the  following
parameterisation:

\begin{equation} 
I_{\rm d}(r_{\rm d})=I_{\rm 0}\, \left[e^{\frac{-r_{\rm d}}{h}}\, \theta \, + \, e^{\frac{-r_{\rm break}\,(h_{\rm out}-h)}{h_{\rm out}\,h}}\, e^{\frac{-r_{\rm d}}{h_{\rm out}}}\,(1-\theta)\right] ,
\label{eqn:disk_trunc} 
\end{equation} 
%
where
\[ \theta =
\begin{cases}
   0  & \qquad {\rm if} \qquad r_{\rm d} > r_{\rm break} \nonumber \\
   1  & \qquad {\rm if} \qquad r_{\rm d} < r_{\rm break},\\
\end{cases}
\]
%
and $r_{\rm  d}$ is the  radius measured in the  Cartesian coordinates
describing  the reference  system of  the disc.   $I_0$, $h$,  $h_{\rm
  out}$, and $r_{\rm break}$ are the central surface brightness, inner
scale-length,  outer  scale-length,  and  break radius  of  the  disc,
respectively. A set of more  elaborate functions to describe the break
of   exponential   discs  has   been   proposed   in  the   literature
\citep[e.g.][]{peng10,erwin15}.   However,  we  opted  for  a  simpler
description  in  order  to  minimise the  number  of  free  parameters
involved in the fitting process.

The projected surface density of a three-dimensional Ferrers ellipsoid
(\citealt{ferrers77},  see   also  \citealt{aguerri09})  is   used  to
describe the SBD of the bar component;

\begin{equation}
I_{\rm bar}(r_{\rm bar})=I_{\rm 0,bar}\left[1-\left(\frac{r_{\rm bar}}{a_{\rm bar}}\right)^2\right]^{n_{\rm bar}+0.5}; 
\qquad {\rm for}\,\, r_{\rm bar} \le a_{\rm bar},
\end{equation}
%
where  $r_{\rm  bar  }$  is  the  radius  measured  in  the  Cartesian
coordinates  describing  the reference  system  of  the bar.   $I_{\rm
  0,bar}$,  $a_{\rm bar}$,  and  $n_{\rm bar}$  represent the  central
surface  brightness,   length,  and   shape  parameter  of   the  bar,
respectively.  Due to  the high degree of degeneracy  that the $n_{\rm
  bar}$ parameter introduces during the fit,  we decided to keep it as
a fixed parameter  during the fitting process. The  default value used
was $n_{\rm bar}=2$ \citep[see also][]{laurikainen05}. Two galaxies in
the sample appeared  to host a nuclear bar after  a careful inspection
of the  2D residuals  and their  SBD was also  fitted using  a Ferrers
profile.

\begin{figure*}
\includegraphics[angle=-90,width=\textwidth]{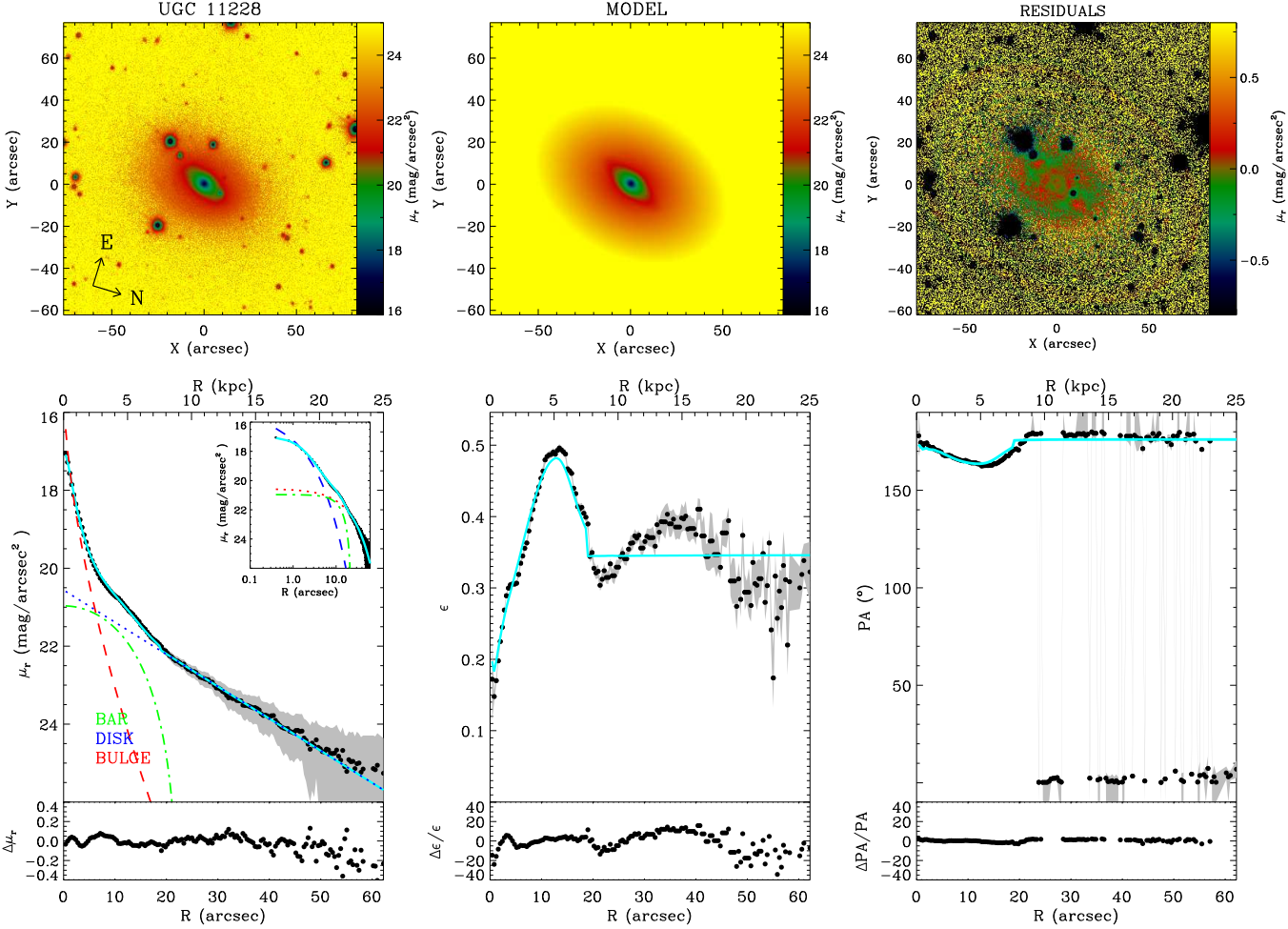}
\caption{Example of a  diagnostic figure used to  determine the number
  of stellar components for each  galaxy. The plot represents the best
  fit using three  components (bulge, bar, and disc)  for the $r-$band
  image of  UGC~11228.  Similar  plots were created  for the  $g-$ and
  $i-$band to  check for consistency.   Top left panel:  galaxy image.
  Top middle panel: best-fitting model of the galaxy image obtained by
  adding  a bulge,  a bar,  and a  disc component.   Top right  panel:
  residual image obtained  by subtracting the best-fit  model from the
  galaxy   image.   Bottom   left   panel:  ellipse-averaged   surface
  brightness radial  profile of the  galaxy (black dots)  and best-fit
  model  (cyan solid  line).   The light  contributions  of the  bulge
  (dashed red line),  disc (dotted blue line),  and bar (dotted-dashed
  green  line)  are shown.   The  upper  inset  shows  a zoom  of  the
  surface-brightness data  and fit  with a  logarithmic scale  for the
  distance  to  the  center  of the  galaxy.   1D  surface  brightness
  residuals  (in  mag/arcsec$^2$  units)   are  shown  in  the  bottom
  sub-panel.  Bottom middle panel:  ellipse-averaged radial profile of
  ellipticity  of the  galaxy (black  dots) and  best-fit model  (cyan
  solid line).  1D  residuals (in percentage) are shown  in the bottom
  sub-panel.  Bottom  right panel: ellipse-averaged radial  profile of
  position angle of  the galaxy (black dots) and  best-fit model (cyan
  solid line).  1D  residuals (in percentage) are shown  in the bottom
  sub-panel. The grey shaded areas  in the bottom panels represent the
  measurement errors  derived from  the {\tt  ellipse IRAF}  task when
  applied to the galaxy image.}
\label{fig:examplefit}
\end{figure*}

To  derive  the photometric  parameters  of  the different  structures
present in a  given galaxy, we iteratively fit a  composite model made
of  a  suitable  combination   of  the  previously  described  stellar
components.    The   actual  fitting   for   each   galaxy  is   human
supervised. This means that the final number of components included in
the fit is based on the judgement of the code-user.  Usually, at least
two different component combinations are tested for each galaxy before
the  best  solution  is  found.   The decision  is  based  on  the  2D
distribution of the intensity residuals and the 1D surface brightness,
ellipticity,    and    position    angle    radial    profiles    (see
Sect.~\ref{sec:multiphot}).    Fig.~\ref{fig:examplefit}    shows   an
example  of the  plots  used to  choose the  final  number of  stellar
structures in UGC~11228.  Upper panels show the 2D SBD for the galaxy,
model, and  residuals, and  the lower panels  represent the  1D radial
profiles of the surface brightness, ellipticity and position angle. In
this particular case, the best fit is achieved using a three-component
model  with   a  bulge,  a   bar,  and  a  single   exponential  disc.
Fig.~\ref{fig:examplefit}  clearly  shows  that  a  break  exponential
profile is  not necessary to reproduce  the SBD of UGC~11228,  but the
bar  is indispensable  to reproduce  the  bump in  the 1D  azimuthally
averaged ellipticity and PA profiles.

An exception to the previous human supervised fitting procedure is the
case of featureless early-type galaxies. If a given galaxy is visually
classified as  early-type (elliptical  or S0/0a) and  other structures
such as  bars or breaks  are not evident in  the image, the  number of
components,  that is,  whether  the  galaxy is  fitted  with a  single
S\'ersic (elliptical) or  a bulge+disc (S0/0a), is  decided through an
automatic criteria. More details about  this procedure can be found in
Sect.~\ref{sec:onevsmulti}.

Increasing the level  of complexity of the galaxy  model, by including
extra components  such as  lenses, ovals,  spiral arms,  or barlenses,
might  improve  the  quality  of  the final  fit,  and  provide  extra
information about the galaxy structure.  However, it comes at the cost
of greatly increasing  the degeneracy on the  final parameters, making
their interpretation  difficult. Therefore, we decided  not to include
any further structures in our analysis.

\subsection{Pre-processing SDSS images}
\label{sec:preprocessing}

In order to  perform the 2D photometric decomposition,  GASP2D needs a
series of input files: (1) a sky-subtracted image of the galaxy; (2) a
mask created to avoid foreground stars, background galaxies, and other
galactic features departing from the  smooth light distribution of the
galaxy; and  (3) the radial  profiles of ellipticity,  position angle,
and surface  brightness.  To compute such  inputs, we make use  of the
fully-calibrated $g-$, $r-$, and $i-$band images from the SDSS-DR7.

\subsubsection{SDSS image sky subtraction, masks, and isophotal analysis}
\label{sec:sdssimages}

\begin{figure}[!h]
\includegraphics[bb=20 320 300 720,width=0.49\textwidth]{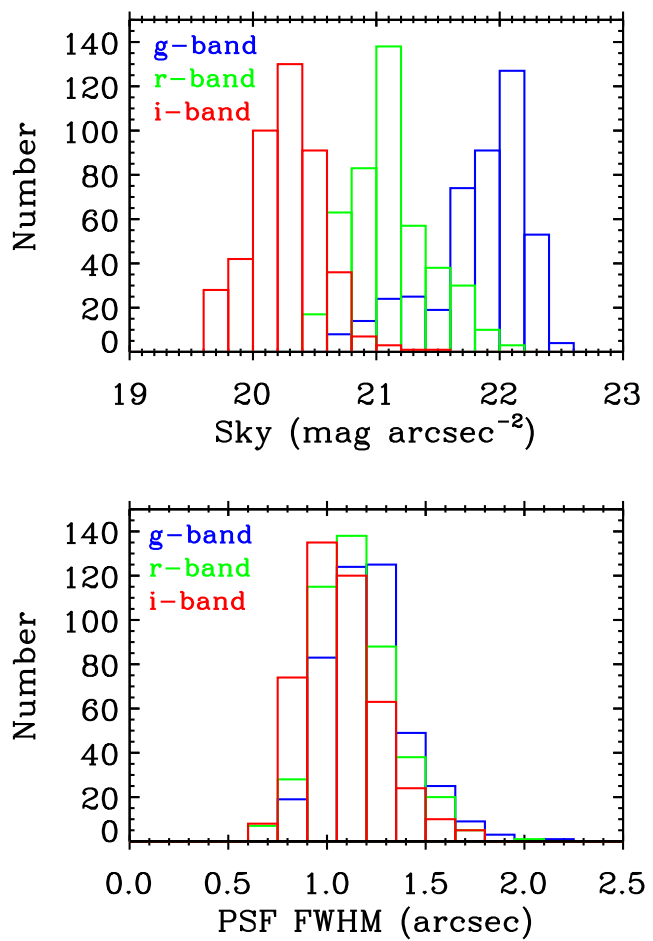}
\caption{Top panel: Distribution of  the average surface brightness of
  the sky subtracted from our galaxy images.  Bottom panel: Distribution
  of the  PSF FWHM  in our  galaxy images.  In  both panels  the blue,
  green, and red  histograms represent the $g-$,  $r-$, and $i-$bands,
  respectively.}
\label{fig:sky_psf}
\end{figure}

The SDSS-DR7 used in this paper,  as well as later SDSS data releases,
provides a measurement  of the sky level (usually the  median value of
every pixel in a frame of $\sim$13.51 $\times$ 9.83 arcmin$^{2}$ after
a  sigma-clipping   is  applied).   This  estimate   has  been  proven
inadequate for  some studies \citep{hydebernardi09},  especially those
focused  on  the analysis  of  the  faintest  parts of  disc  galaxies
\citep[][]{pohlentrujillo06}. Since  we intend  to provide  a detailed
inventory  and characterisation  of  galaxy structures  in the  CALIFA
survey, an improved sky subtraction  procedure was needed. The details
of this procedure  are explained in a  companion paper (M\'endez-Abreu
et al., submitted).  For the sake  of clarity we will briefly describe
here its  main characteristics. First, we  created an ad hoc  mask for
each   galaxy  frame   using  both   the  automatic   code  SExtractor
\citep{bertinarnouts96}  and a  manually-built mask  to include  small
features  that SExtractor  might have  missed. Then,  we use  the {\tt
  ellipse  IRAF}\footnote{{\tt IRAF}  is distributed  by the  National
  Optical Astronomy Observatory, which  is operated by the Association
  of Universities  for Research in Astronomy  (AURA) under cooperative
  agreement with the National Science  Foundation.} task to obtain the
1D surface-brightness  profile using a fixed  ellipticity and position
angle matching the outermost disc  isophotes.  The actual value of the
sky  is then  computed by  averaging the  region with  a flat  surface
brightness at a radius where neither  the galaxy nor other sources are
affecting  the  flux.  Finally,  this  value  is subtracted  from  the
corresponding science  frame to produce the  sky-subtracted image used
in this paper.  The distribution of  the derived values of the surface
brightness of the sky for the  images of the sample galaxies are shown
in the top panel of Fig.~\ref{fig:sky_psf}.  We find typical values of
21.8$\pm$1.2, 21.1$\pm$1.2, and  20.2$\pm$1.2 mag/arcsec$^2$ for $g-$,
$r-$, and $i-$band, respectively.

The final isophotal profiles used in this study were derived using the
sky-subtracted  images.   We  ran  {\tt ellipse}  again  allowing  the
isophotes to  change the  values of  $\epsilon$ and  PA to  follow the
galaxy  morphology.   The masks  created  during  the sky  subtraction
process were used to derive these 1D profiles and in the actual 2D fit
using GASP2D.  Examples of the  surface brightness, $\epsilon$, and PA
1D     azimuthally     averaged      profiles     are     shown     in
Fig.~\ref{fig:examplefit}.

\subsection{PSF}
\label{sec:psf}

It  is well-known  that accurate  measurements  of the  image PSF  are
critical  for recovery  of  the structural  parameters  of the  galaxy
central  components.   In  this  study   this  mainly  refers  to  the
properties of bulges and NPSs. \citet{mendezabreu08} found that errors
of $\sim$2\% in the PSF FWHM can produce variations of up to $10\%$ in
the  $r_e$ and  $n$  bulge  parameters.  Similarly,  \citet{gadotti08}
showed  that  the structural  properties  of  bulges can  be  reliably
retrieved provided  that $r_e$ is larger  than $\sim 80\%$ of  the PSF
half   width   half  maximum   (see   also   Costantin  et   al.,   in
prep). Therefore,  a careful  analysis of  the SDSS  PSF is  needed to
perform a robust photometric decomposition.

We model the SDSS PSF  with a Moffat function (see Eq.~\ref{eqn:psf}).
This parameterisation of  the PSF has been extensively  studied in the
literature   \citep[e.g.][]{trujillo01}  and   provides  an   accurate
representation of  the SDSS PSF.  For  each galaxy image, a  set of at
least  five non-saturated  stars were  fitted with  a Moffat  function
using the {\tt  IRAF} task {\tt imexam}.  The mean  values of the FWHM
for   our  galaxy   sample  are   shown   in  the   bottom  panel   of
Fig.~\ref{fig:sky_psf}.   We  find   typical  values  of  1.2$\pm$0.2,
1.1$\pm$0.2, and 1.1$\pm$0.2 arcsecs for the $g-$, $r-$, and $i-$band,
respectively.  GASP2D uses the measured  PSF as kernel to be convolved
with  the model  galaxy  image before  computing  the $\chi^2$.   This
process  is  repeated in  each  iteration  of the  Levenberg-Marquardt
minimisation  process  so  that  the  final  best-fit  parameters  are
seeing-corrected.

\section{2D photometric decomposition analysis: multi-component fitting}
\label{sec:multiphot}

This section describes the main procedures  we follow to carry out the
photometric  decomposition.  We  separate our  sample into  early-type
(127 galaxies) and spiral galaxies (277 galaxies) since for the former
an  automatic  methodology is  used  to  find  the optimal  number  of
components  to obtain  the best  fit whereas  for the  latter a  human
supervised approach  is used.  Table \ref{tab:tabfits}  shows the best
fit parameters  obtained for some example  galaxies covering different
combinations of structural  components.  The full version  of the table
for the entire galaxy sample is available on-line.

\begin{table*}[!ht]
\caption{Structural parameters of the sample galaxies in the $r-$band}   
\label{tab:tabfits}    
\centering            
\begin{tabular}{l c c c c c c}      
\hline\hline   
Galaxy           & NGC0155           & NGC0001            & NGC0160          & UGC00036          & NGC7819           &   NGC0941           \\  
\hline
$\mu_e$          &  22.6$\pm$ 0.1     &  19.8$\pm$ 0.2    & 20.0$\pm$0.1     &  19.1$\pm$ 0.3    &  19.7$\pm$ 0.2    &     --               \\  
$r_{\rm e}$       &  28.6$\pm$ 1.7     &  4.6$\pm$ 0.5     & 6.2$\pm$0.5      &  1.4$\pm$ 0.3     &  2.4$\pm$ 0.3     &  --                  \\  
$n$              &  5.0$\pm$ 0.2      &  2.8$\pm$ 0.2     & 2.6$\pm$0.1      &  1.9$\pm$ 0.3     &  1.1$\pm$ 0.1     &  --                  \\  
$q_{\rm bulge}$    &  0.76$\pm$ 0.01    & 0.80$\pm$ 0.02    & 0.71$\pm$0.02   & 0.71$\pm$ 0.05     & 0.65$\pm$ 0.04    &  --                  \\ 
PA$_{\rm bulge}$   &  174.1$\pm$ 0.6    &  128.2$\pm$ 3.1   & 49.5$\pm$4.2    &  24.5$\pm$ 5.6     &  88.7$\pm$ 5.2    &       --             \\ 
$B/T$            &    1.0             &   0.46            & 0.41             &   0.11            &   0.12            &      --               \\ 
 &  &  & & &   \\
$\mu_0$           &     --            &   20.7$\pm$ 0.1   & 21.55$\pm$0.03   &   19.5$\pm$ 0.1   &   22.1$\pm$ 0.1   &      20.68$\pm$ 0.01  \\ 
$h$               &     --            &   14.9$\pm$ 0.9   & 47.4$\pm$0.9     &   9.5$\pm$ 0.4    &   44.7$\pm$ 2.5   &      20.2$\pm$ 0.4    \\  
$r_{\rm break}$    &     --            &       --           & 48.6$\pm$1.0     &        --         &  46.9$\pm$ 4.4    &           --          \\   
$h_{\rm out}$      &     --            &      --           &  14.1$\pm$0.7    &        --          &  12.1$\pm$ 1.5    &           --          \\  
$q_{\rm disc}$     &     --            & 0.62$\pm$ 0.01    &  0.505$\pm$0.9    & 0.45$\pm$ 0.01    & 0.53$\pm$ 0.01    &  0.84$\pm$ 0.01        \\   
PA$_{\rm disc}$    &     --            &    96.5$\pm$ 1.1  & 47.5$\pm$0.2      &    17.9$\pm$ 0.7  &    102.0$\pm$ 0.5 &        166.0$\pm$ 0.7  \\  
$D/T$             &    --             &   0.54            &  0.59            &   0.79             &   0.75            &      0.996             \\ 
 &  &  & & &   \\
$\mu_{\rm 0, bar}$  &      --           &      --          &          --       &   20.5$\pm$ 0.1    &   22.0$\pm$ 0.1   &        --             \\   
$a_{\rm bar}$      &      --            &      --          &         --        &   12.4$\pm$ 0.3    &   39.1$\pm$ 1.2   &       --              \\  
$q_{\rm bar}$      &      --            &      --          &        --         & 0.63$\pm$ 0.01     & 0.32$\pm$ 0.01    &      --               \\   
PA$_{\rm bar}$     &      --            &      --          &       --          &    134.0$\pm$ 0.4  &    59.4$\pm$ 0.4  &        --             \\
$Bar/T$           &      --            &      --          &       --          &   0.10             &   0.13            &      --               \\ 
 &  &  & & &   \\
$\mu_{\rm 0, NPS}$  &     --             &     --          &       --           &      --            &       --          &       18.2$\pm$ 0.1  \\   
$NPS/T$            &     --            &      --         &       --           &      --            &       --          &        0.004          \\ 
 &  &  & & &   \\
$Flag$            &    1,a            &    1,a           &    1,a             &    2,a             &    2,c            &    2,c                 \\ 
\hline                                 
\end{tabular}
\tablefoot{Best-fit  parameters  for  a   subsample  of  six  galaxies
  modelled with  a different combination  of structures. From  left to
  right:  single S\'ersic,  bulge+disc, bulge+disc  (including break),
  bulge+disc+bar,  bulge+disc  (including  break)+bar,  and  NPS+disc.
  Each column represents  the best-fit parameters for  a given galaxy.
  From top  to bottom: bulge parameters  (effective surface brightness
  $\mu_e$,  effective radius  $r_{\rm  e}$, S\'ersic  index $n$,  axis
  ratio  $q_{\rm   bulge}$,  position  angle  PA$_{\rm   bulge}$,  and
  bulge-to-total  luminosity ratio  $B/T$),  disc parameters  (central
  surface  brightness $\mu_0$,  inner scale  length $h$,  break radius
  $r_{\rm  break}$,  outer  scale  length $h_{\rm  out}$,  axis  ratio
  $q_{\rm disc}$,  position angle  PA$_{\rm disc}$,  and disc-to-total
  luminosity ratio $D/T$), bar  parameters (central surface brightness
  $\mu_{\rm 0,  bar}$, bar  radius $a_{\rm  bar}$, axis  ratio $q_{\rm
    bar}$, position angle PA$_{\rm  bar}$, and bar-to-total luminosity
  ratio $Bar/T$), NPS parameters (central surface brightness $\mu_{\rm
    0,  NPS}$  and NPS-to-total  luminosity  ratio  $NPS/T$), and  the
  visual quality flag explained  in Sect.~\ref{sec:quality}.  Each row
  shows  the best  fitting values  of the  given parameters  and their
  associated  error (see  Sect.~\ref{sec:mock}).  Surface  brightness,
  radii, and  position angles  are given  in units  of mag/arcsec$^2$,
  arcsec, and  degrees measured  counterclockwise from North  to East,
  respectively.  When a  given structure is not present  in the model,
  its corresponding  parameters are left  empty.  The full  version of
  the table is  available on-line.  Similar tables  are also available
  for the $g-$  and $i-$bands.  The parameters in this  table have not
  been   corrected   for   galactic   extinction,   K-correction,   or
  cosmological dimming.}
\end{table*}

\subsection{Multi-wavelength fitting process}
\label{sec:multiwav}

The photometric  properties of  the different stellar  structures were
derived independently  for the  three SDSS bands  ($g$, $r$,  and $i$)
used in this paper. This means  that the structural parameters are not
limited and/or  tied between the  different band images.   However, in
order to  avoid discordant  results we decided  to relate  the initial
conditions required  for the fit in  the three bands. In  its standard
configuration,  GASP2D finds  the best  set of  initial conditions  to
initialise the non-linear fit in  an automatic way.  This procedure is
described  in \citet{mendezabreu08}  and is  mainly  based on  the
analysis    of     the    1D    radial    profiles     explained    in
Sect.~\ref{sec:sdssimages}.  In  this study, the galaxy  images in the
$r-$band (the intermediate wavelength band) were fitted following this
standard  procedure  with automatic  initial  conditions  or, in  some
cases, fine-tuning them after a  visual inspection.  Once a successful
fit is achieved, the best-fit parameters  in the $r-$band were used as
initial  conditions for  the $g-$  and $i-$bands.   We find  that this
strategy generally  produces consistent results among  different bands
without constraining the final photometric parameters.

\subsection{One-component vs. multi-component fits of early-type galaxies}
\label{sec:onevsmulti}

The  photometric  properties   of  early-type  galaxies,  encompassing
elliptical and  lenticular galaxies, have been  extensively studied in
the    literature   \citep[see][for    reviews]{kormendy09,aguerri12}.
However, the problem of identifying  whether a stellar disc is present
or  not  in these  smooth  and  featureless  galaxies is  still  under
discussion  \citep{gomes16}.   From  a   photometric  point  of  view,
elliptical galaxies  are stellar  systems well  described by  a single
S\'ersic profile.  On the other hand, the simplest description of a S0
galaxy  is  a  two-component  model,   that  is,  a  S\'ersic  profile
describing the  SBD of the  bulge and a pure  exponential representing
the outer  disc.  It is  worth noting  that this definition  is purely
photometric and, therefore,  is not directly related  to the dynamical
status of the galaxies \citep[see][]{emsellem11}.

From a  photometric point  of view,  advances in  the field  have been
driven  by   the  application  of  statistical   techniques  on  model
selection,   such   as   the    F-test   \citep[see][for   a   similar
  application]{simard11},    the    Bayesian    inference    criterion
\citep[$BIC$;][]{schwarz78},  or  the   Akaike  information  criterion
\citep[$AIC$;][]{akaike74}.   These   techniques  work  by   adding  a
penalisation to  the standard  $\chi^2$ accounting  for the  number of
free  parameters included  in the  fit.  Thus,  these criteria  can be
applied to determine whether or not adding an extra component (i.e. an
outer    disc)   would    statistically   improve    the   best    fit
\citep[e.g.][]{simard11,head14}.  On  the other hand,  these automatic
criteria do  not account  for the  possibility that,  even if  a given
model   is   statistically   preferred,    its   solution   might   be
unphysical. Therefore,  to provide  the best  mathematical fit  with a
physical meaning,  some authors  have proposed the  use of  a `logical
filter'  \citep{allen06}.  In  M\'endez-Abreu et  al.  (submitted)  we
combined   the   two   aforementioned   approaches   to   assess   the
appropriateness of  different model  decompositions and decide  when a
complex  model,  bulge+disc,  is  preferred  over  a  single  S\'ersic
profile.   The  main  features  of   the  logical  filter  consist  of
classifying as  ellipticals those galaxies where  two-component models
produce either:  i) a  large $B/T$  ($B/T > $0.9),  ii) a  large $r_e$
($r_e >  1.676 \times h$),  iii) an even number  (0 or 2)  of crossing
points between the surface-brightness distribution of the disc and the
bulge,  or  iv)  the  previous crossing  point  happening  before  one
effective radius of the disc ($r_{\rm cross} < 1.676 \times h$).  Note
that   galaxies  hosting   stellar   bars  are   relatively  easy   to
identify. Since  bars can be  used to  detect the presence  of stellar
discs    \citep{mendezabreu10,mendezabreu12},   they    are   directly
classified  as S0  without  the  need to  satisfy  either the  logical
filtering or BIC criteria.

\begin{table}[!t]
\caption{Schematic of the sample selection process of early-type galaxies}
\begin{center}
\begin{tabular}{|c|c|c|c|}
\hline
CALIFA VISUAL       &  L. Filter   & L. Filter + $\Delta BIC$   &  FINAL             \\
  (1)               &  (2)  &     (3)             &   (4)              \\
\hline
    \multirow{2}{*}{85 E}& 35 E                   & 35 E          &       \\ \cline{2-3}
                         & \multirow{2}{*}{50 S0} & 30 B/BD       &       \\
                         &                        & 20 S0         & 41 E  \\
    \cline{1-3}
    \multirow{2}{*}{42 S0}& 6 E                   & 6 E           & 36 B/BD  \\\cline{2-3}
                          & \multirow{2}{*}{36 S0}& 6 B/BD        & 50 S0 \\
                          &                       & 30 S0         &       \\
    \hline
\end{tabular}
\tablefoot{E-Elliptical, S0-Lenticular, B/BD-Elliptical or Lenticular.
  (1) Number of  galaxies using the CALIFA  visual classification; (2)
  number  of galaxies  after the  logical filtering  (L. Filter);  (3)
  number   of  galaxies   after  the   logical  filtering   and  $BIC$
  classification; (4) final classification used in this study.}
\end{center}
\label{tab:ellipvss0}
\end{table}

We therefore applied the  two-step process described in M\'endez-Abreu
et  al.  ({submitted})  to  our  sample of  127  early type  galaxies.
Table~\ref{tab:ellipvss0} shows the outcome  of our analysis. Galaxies
were first divided into ellipticals and S0 based on the outcome of the
logical filter.   Then, for  those galaxies where  the two  models are
compatible, the $BIC$  analysis is performed to  discern whether there
is statistical evidence against one of  the models or not. Still, some
galaxies remain  with unclear classification, equally  compatible with
being a single S\'ersic or a two-component galaxy. This last sample is
labelled in the tables as B/BD morphology and highlights the intrinsic
difficulties  of   separating  ellipticals  from  S0   galaxies  using
photometric data.  The best-fit obtained from both the single S\'ersic
and the bulge+disc  is provided for these galaxies to  allow the users
to decide which decomposition is more suitable for their science case.
We end up  with a final sample of 41  ellipticals, 50 lenticulars, and
36 galaxies with B/BD morphology.

\subsection{Multi-component analysis of spiral galaxies}
\label{sec:multiphotspiral}

The final  decomposition of our  CALIFA sample of spiral  galaxies was
done  using  a  {human-supervised}   approach  fitting  up  to  three
components:  bulge/NPS, disc  (including  break), and  bar.  For  each
galaxy, a given combination of these structures provides the best fit.

Small  bulges with  sizes comparable  to the  SDSS PSF  can lead  to
erroneous fits, usually producing extreme values of the S\'ersic index
($n>7$). Although one of our goals is to produce reliable estimates of
the bulge parameters, high values of $n$ might also have a significant
impact on other  components' parameters. Therefore, we  also fit these
galaxies using a NPS instead  of the S\'ersic parameterisation for the
bulge.   A visual  inspection  of  the 2D  residual  is  then used  to
determine which component is preferred.  As a consequence, a bulge and
a NPS  would never be  fitted simultaneously  in the same  galaxy, and
galaxies  better represented  by a  NPS  should not  be understood  as
necessarily hosting a nuclear star cluster or AGN.  GASP2D also allows
for  different behaviours  of  the outer  disc component  (exponential
profile, type  I; and  broken profiles,  type II  and type  III).  Our
CALIFA  sample  was fitted  using  13  different combinations  of  the
previous   structures   (see   Table~\ref{tab:tabcomponents}).    This
demonstrates  the  morphological  variety  of  our  galaxies  and  the
necessity   of   performing   accurate   multi-component   photometric
decompositions.

In  Sect~\ref{sec:onevsmulti}  we  described the  particular  case  of
fitting early-type galaxies.  The detection of stellar discs in spiral
galaxies  is more  direct  than  for early-type  galaxies  due to  the
presence  of  star  formation   following  the  characteristic  spiral
pattern.    The  strategy   followed   to   perform  the   photometric
decomposition  of   a  spiral  galaxy  starts   with  a  two-component
bulge+disc fit.   Right after, or  from the  start if its  presence is
readily apparent in  the galaxy image, we check for  the addition of a
bar component (see Sect~\ref{sec:bars}).  If the bar is not obvious in
the galaxy image, both the  bulge+disc and bulge+disc+bar fittings are
performed. Then, a careful visual analysis  of the 2D SBD residual, as
well of the  1D ellipticity and PA profiles (see  Aguerri et al.  2009
for a description of bar identification through the galaxy ellipticity
and PA) is carried out to reveal  whether a bar is actually present or
not. It is worth noting that the detection of bars is limited to those
central structures with  an axis ratio $q_{\rm bar} <$  0.7.  The next
step  is  to check  for  the  presence of  broken  discs  and, if  the
following conditions are fulfilled, include  them in the fit: i) after
a careful revision of the 2D  SBD residual we confirm potential breaks
are not  misidentified with  spiral arms and/or  outer rings,  and ii)
they appear  at a SB  level $\mu_i<24$  mag/arcsec$^2$ so they  can be
robustly     measured      with     the     SDSS      imaging     (see
Sect.~\ref{sec:breaklimits}).

The  different combinations  of structures  used in  our final  galaxy
sample are  shown in Table~\ref{tab:tabcomponents}.  In  the following
sections  we  will describe  the  incidence  of the  different  galaxy
structures in the CALIFA sample.

\begin{table}[!t]
\caption{Different   combinations   of   structures  present   in   the
  photometric decomposition of our sample galaxies}
\label{tab:tabcomponents}    
\centering            
\begin{tabular}{l c}      
\hline\hline              
Structure      &  Number       \\  
   (1)         &    (2)        \\
\hline
B              &   43          \\ 
D              &   6           \\ 
BD             &   74          \\
ND             &   12          \\
BD$_{\rm br}$   &   67          \\
D$_{\rm br}$    &   4           \\
BDBar          &   88          \\
DBar           &   7           \\ 
NDBar          &   10           \\
BD$_{\rm br}$Bar  &   47        \\
D$_{\rm br}$Bar   &   3         \\
ND$_{\rm br}$Bar  &   5         \\
BDBarNBar        &   2         \\
B/BD           &   36          \\
\hline                                 
\end{tabular}
\tablefoot{(1) Type of structure: B -  single S\'ersic, D - pure disc,
  BD -  bulge+disc, ND  - nuclear point  source+disc, BD$_{\rm  br}$ -
  bulge+disc with break, D$_{\rm br}$ -  pure disc with break, BDBar -
  bulge+disc+bar,   DBar   -   disc+bar,   NDBar   -   nuclear   point
  source+disc+bar,  BD$_{\rm  br}$Bar  -  bulge+disc  with  break+bar,
  D$_{\rm br}$Bar  - disc with  break+bar, ND$_{\rm br}$Bar  - nuclear
  point source+disc with break+bar, BDBarNBar - bulge+disc+bar+nuclear
  bar,  B/BD -  either single  S\'ersic or  bulge+disc; (2)  number of
  galaxies.}
\end{table}

\subsubsection{Barred galaxies}
\label{sec:bars}

The inclusion of the bar SBD in the photometric decomposition has been
proven to  be critical in  order to recover accurate  bulge parameters
\citep[e.g.][]{aguerri05,laurikainen05}.  Several  studies have shown
that  both the  S\'ersic index  and the  $B/T$ ratio  can artificially
increase  if  the  bar  is  not properly  accounted  for  in  the  fit
\citep{gadotti08,salo15}.

\begin{figure*}
\includegraphics[bb=54 350 350 600]{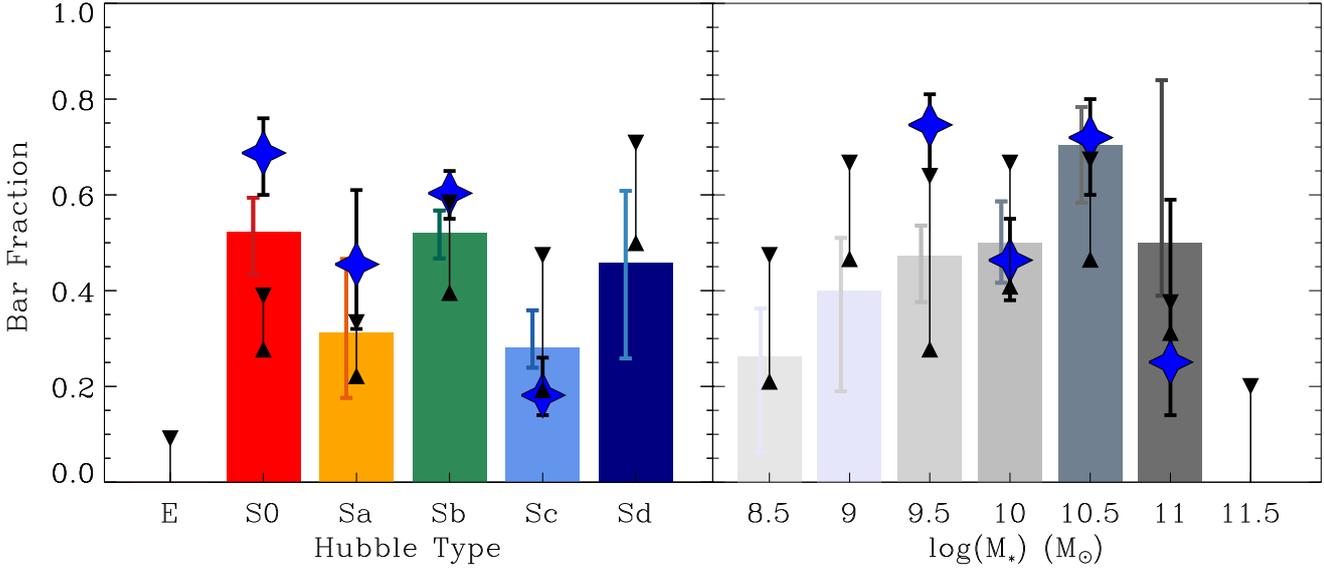}
\caption{Distribution of the bar fraction  as a function of the Hubble
  type  (left panel)  and  stellar mass  (right  panel).  Colour  bars
  represent  the photometric  bar  fraction derived  from this  study.
  Lines with darker colours show the 1$\sigma$ error.  Blue stars show
  the volume corrected  bar fractions using only  those galaxies drawn
  from  the mother  sample  and with  luminosities  within the  CALIFA
  completeness limits  (285 galaxies).  The black  lines represent the
  visual classification  from \citet{walcher14} with the  upper limits
  including both  weak (AB) and  strong (B)  bars and the  lower limit
  accounting   only   for   strong   bars.   Note   that   since   the
  \citet{walcher14} Hubble type and bar identification are the average
  among different  independent classifications, some  ellipticals were
  classified as weakly barred.  Bins  with less than five galaxies are
  not shown.}
\label{fig:bars}
\end{figure*}

It  is  worth noting  that  the  inclusion of  a  stellar  bar in  our
photometric decomposition  is independent of the  visual morphological
classification \citep[see][for details]{walcher14}.  Therefore, we can
quantify the impact of different bar classification methods on studies
of the galaxy bar fraction. Fig.~\ref{fig:bars} shows the distribution
of  the bar  fraction using  either the  visual classification  or the
photometric  decomposition method  with respect  to the  galaxy Hubble
type  and stellar  mass.  In  addition,  we have  included the  volume
corrected  bar  fractions  for the  photometric  decomposition  method
applied to those galaxies extracted from the CALIFA mother sample (see
Sect.~\ref{sec:sampleselection}). A summary of the result is presented
in Table~\ref{tab:barfraction}.  In general,  we find a good agreement
in  the  observed   bar  fractions  obtained  using   the  visual  and
photometric  decomposition  method.   The uncertainty  in  the  visual
identification of a bar is enough  to account for the differences with
respect to the photometric decompositions for all cases except for the
galaxies classified as S0 and Sd. In  the case of the S0, not only the
presence  of a  bar but  also the  morphological classification  as S0
itself  depends on  the method  (see Sect.~\ref{sec:onevsmulti}).   We
attribute  the differences  in the  Sd  galaxies to  the small  number
statistics (only 24 galaxies are  Sd).  The volume corrected fractions
can  deviate substantially  from  the estimates  from the  photometric
decomposition method.
  
\begin{table}[!t]
\caption{Volume corrected bar fraction distributions in our galaxy sample}
\begin{center}
\begin{tabular}{l c}
\hline
Hubble Type  &  Bar fraction\\
\hline
S0   & 68.7\%$\pm$7.2\% \\
Sa   & 45.5\%$\pm$15.5\%  \\
Sb   & 60.3\%$\pm$4.7\%  \\
Sc   & 18.2\%$\pm$7.8\% \\
Sd   &  --          \\
\hline
Stellar Mass & Bar fraction\\
\hline
$9 < {\rm log(M_{\star}/M_{\sun})}>9.5$   &  --           \\
$9.5 < {\rm log(M_{\star}/M_{\sun})}>10$  & 74.6\%$\pm$6.4\%  \\
$10 < {\rm log(M_{\star}/M_{\sun})}>10.5$ & 46.4\%$\pm$8.6\% \\
$10.5 < {\rm log(M_{\star}/M_{\sun})}>11$ & 72.0\%$\pm$8.0\% \\
$11 < {\rm log(M_{\star}/M_{\sun})}>11.5$ & 25.1\%$\pm$22.1\% \\
\hline
\end{tabular}
\end{center}
\label{tab:barfraction}
\end{table}

The  influence of  galaxy  morphology  on the  bar  fraction has  been
extensively   discussed   in   the   literature   with   contradictory
results. Several authors have claimed  that the bar fraction increases
towards  early-type  galaxies  \citep{masters11,masters12,lee12}  with
this  trend being  consistent with  some models  of bar  formation and
evolution  \citep{athanassoula13}.  Nevertheless,  other studies  have
found      the      opposite      trend     with      Hubble      type
\citep[][]{laurikainen07,barazza08,aguerri09,buta15}.              Our
volume-corrected bar  fraction is relatively constant  for early-types
(S0,  Sa,  and  Sb)  but   it  dramatically  drops  for  Sc  galaxies.
Unfortunately,  small  number  statistics  for the  Sd  type  preclude
further analysis of that bin, so we cannot confirm the drop in the bar
fraction for  all late-type  galaxies. We  find a  mean value  for the
volume-corrected bar fraction of 57\%.  This number is similar to that
obtained   using   the  observed   sample   (54\%)   and  the   visual
classification  (51\%).  Recently,  \citet{buta15} found  a lower  bar
fraction  in  early-type  galaxies  ($\sim 56\%$)  than  in  late-type
galaxies  ($80\%$) using  a visual  classification of  the S4G  galaxy
sample  \citep{sheth10}.   They  suggested  that  the  different  mass
distribution of  galaxies with both early-  and late-type morphologies
could, however, be driving this result.

In  fact,  the  previous,  apparently  contradictory  results  can  be
reconciled when the galaxy stellar mass is accounted for in the sample
selection  \citep{nairabraham10}.  The  incidence  of  bars in  galaxy
discs is a strong function of stellar mass with a maximum bar fraction
at  $M_{\star}\sim2\times10^9  {\rm   M_{\sun}}$  for  field  galaxies
\citep{mendezabreu10,mendezabreu12}.   The volume  corrected fractions
show an  increase of  the bar fraction  towards lower  stellar masses.
This   trend   is   in   good    agreement   with   the   results   of
\citet{mendezabreu12} although  the CALIFA  sample is not  complete at
$M_{\star} \sim 10^9 M_{\sun}$ where the bar incidence is predicted to
be highest. The observed bar fraction shows a different picture, being
nearly  independent of  stellar mass  within the  errors for  both the
visual and photometric decomposition method.

An interesting sub-sample of barred  systems contains nested bars, that is,
galaxies hosting both a large-scale  bar (described previously) and an
inner, secondary    bar     embedded     in     the    main     one
\citep{erwinsparke02,corsini07,delorenzocaceres12,delorenzocaceres13}.
Whilst they are not  the main focus of this paper,  they are thought to
be present  in $\sim$30\%  of barred  galaxies \citep{laine02,erwin04}
and    might  have  an   impact  on  the  bulge   parameters  (de
Lorenzo-C\'aceres et al.  {\it in prep}). Therefore, when necessary we
included  a  secondary bar  (using  another  Ferrers profile)  in  our
photometric decomposition.  We found only two galaxies with nested bars;
NGC~0023 and NGC~7716,  which represent a much  lower fraction (1.5\%)
with respect to previous findings.  Angular resolution likely explains
these differences. The sample  presented in \citet{erwin04} is located
at an average distance of 29  Mpc (0.144 kpc/arcsec), with some of the
inner bars detected  by using imaging from the  Hubble Space Telescope
Wide Field Planetary  Camera 2 (HST/WFPC2) with a  typical 0.07 arcsec
resolution.   On the  other hand,  our  sample of  barred galaxies  is
located at 72 Mpc (0.346 kpc/arcsec)  limited to the $\sim 1.1$ arcsec
SDSS resolution.   Thus, in our  sample a  low double bar  fraction is
expected.

\begin{figure*}[!ht]
\includegraphics[bb=54 345 688 570,width=1.15\textwidth]{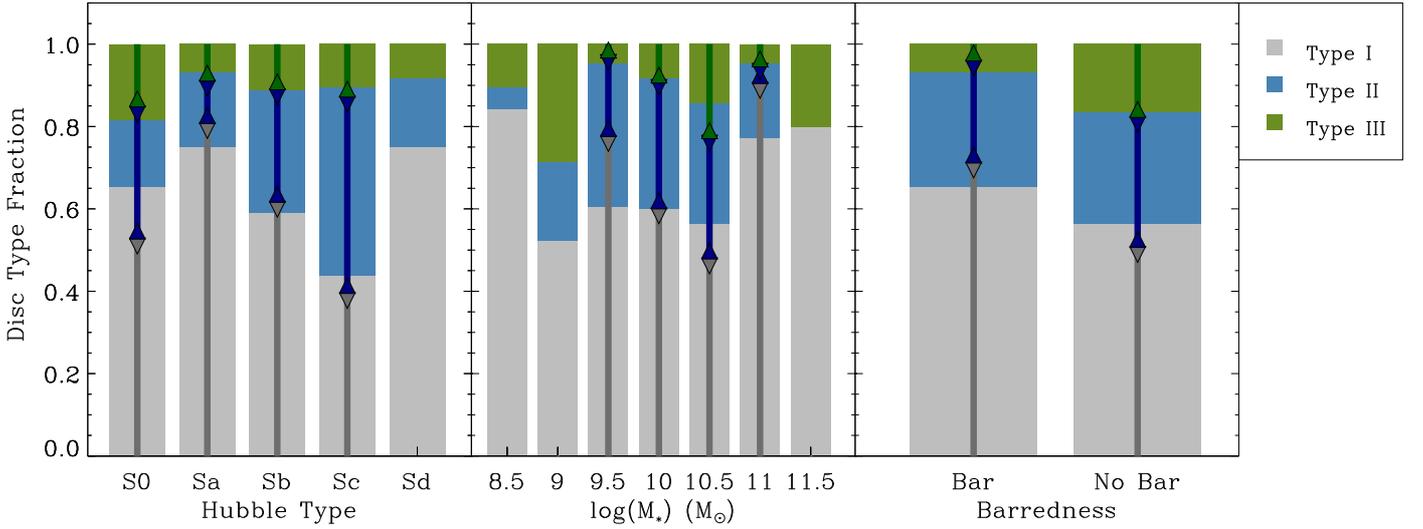}
\caption{Distribution  of the  three different  disc profiles  used in
  this  study with  Hubble  type (left  panel),  stellar mass  (middle
  panel), and presence of a  bar (right panel).  Colour bars represent
  our observed fractions for Type I (grey), Type II (blue), and Type
  III (green).   Lines with darker  colours show the  volume-corrected
  fractions.  Bins with less than five galaxies are not represented.}
\label{fig:breaks}
\end{figure*}

\subsubsection{Breaks in disc galaxies}
\label{sec:breaks}
Galactic  discs  with  non-purely exponential  profiles  represent  an
important fraction of the discs in the Local Universe and are known to
appear   in    galaxies   independently    of   their    Hubble   type
\citep[e.g.][]{erwin08,gutierrez11,marino16}.    The    ubiquity   of
stellar discs with  broken profiles is manifest even  at high redshift
\citep{perez04,trujillopohlen05,azzollini08}, suggesting that they are
key features in understanding how galaxies form and evolve.

From the photometric decomposition point of view, broken profiles have
an impact  on the properties  of the remaining components  included in
the fit. Therefore, a complete  and robust analysis of the photometric
properties of galaxies  must include the possibility  of disc galaxies
displaying   broken  profiles.    Previous   studies   of  the   light
distribution in  the galaxy  outskirts have mainly  been based  on the
analysis of  1D azimuthally averaged  profiles, and thus  are not
representative  of  the  two-dimensional  nature  of  the  problem  of
galaxies. GASP2D is able to perform a multi-component decomposition of
the  galactic light  including  broken  profiles and  this  work is  a
pioneering study on the incidence  of broken profiles in disc galaxies
based on the 2D approach.

\begin{table}[!t]
\caption{Volume corrected disc type distributions in our galaxy sample}
\begin{center}
\begin{tabular}{l c c c}
\hline
Hubble Type  &  Type I & Type II  & Type III\\
\hline
S0   & 52.8\% & 32.1\% & 15.1\%  \\
Sa   & 80.8\% & 10.3\%  & 8.9\%   \\
Sb   & 61.7\% & 27.3\% & 11.0\%   \\
Sc   & 39.6\% & 47.7\% & 12.7\% \\
Sd   &  --  &  --  &   --  \\
\hline
Stellar Mass &  Type I & Type II  & Type III\\
\hline
$9 < {\rm log(M_{\star}/M_{\sun})}>9.5$   & --   & --   & -- \\
$9.5 < {\rm log(M_{\star}/M_{\sun})}>10$  & 77.7\% & 21.7\% & 0.6\%\\
$10 < {\rm log(M_{\star}/M_{\sun})}>10.5$ & 60.1\% & 33.1\% & 6.7\%\\
$10.5 < {\rm log(M_{\star}/M_{\sun})}>11$ & 48.0\% & 31.8\% & 20.2\%\\
$11 < {\rm log(M_{\star}/M_{\sun})}>11.5$ & 90.6\% & 6.6\%  & 2.8\%\\
\hline
Barredness  &  Bar fraction \\
\hline
Bar                & 71.2\%  & 24.8\% & 3.9\%  \\
No Bar             & 50.8\%  & 31.6\% & 17.7\% \\
\hline
\end{tabular}
\end{center}
\label{tab:discfraction}
\end{table}

Figure \ref{fig:breaks}  shows the distribution of  the different disc
profiles as a function of the  Hubble type, stellar mass, and presence
of a bar.  We  find that 62\%, 28\%, and 10\%  of our volume corrected
disc sample is best represented with a  Type I, Type II, and Type III,
respectively. This represents a significantly lower fraction of broken
discs when  compared to 1D  analyses present in either  the literature
\citep[e.g.][]{gutierrez11,laine14}     or    the     CALIFA    sample
\citep{marino16}.  In  Ruiz-Lara et  al.  (in  prep.)  we  performed a
detailed analysis  to understand these differences  and concluded that
the discrepancy is mainly caused by our new 2D approach instead of due
to   selection   effects  in   the   galaxy   samples.   In   general,
multi-component 2D  decompositions such as  the one performed  in this
paper  build the  galaxy model  as a  combination of  different galaxy
structures  that  contribute  differently   to  the  total  luminosity
depending on the galactocentric radius.   This is important in regions
with a  high overlap of structures  such as the area  where the bulge,
disc, and bar coexist.  On the  contrary, most of the previous studies
using 1D  profiles used pre-defined regions  of the surface-brightness
profile  where a  single exponential  is fitted,  without taking  into
account   any  superposition   with  other   galaxy  components   (see
\citealt{munozmateos13} for an estimation  of errors associated to the
pre-defined  disc regions).   This  different methodology  leads to  a
higher fraction of broken profiles  in 1D analysis (in particular Type
II), since it is straightforward to accommodate piecewise exponentials
to different sections of the profile, but it is not straightforward to
associate these piecewise  exponentials with an actual  change in disc
structure. At  the same time, Type  III breaks usually occur  at lower
surface brightness  than Type II \citep{pohlentrujillo06}  so they are
intrinsically  more  difficult  to identify.   In  2D  multi-component
analysis  the addition  of  this  new structure  (i.e.   two new  free
parameters $r_{\rm break}$  and $h_{out}$) is not  always justified in
statistical  terms   (see  Sect.~\ref{sec:breaklimits}),   whereas  1D
analysis methods  can easily  deal with  fitting a  pre-defined region
even if at low SB.

\begin{figure*}[!ht]
\includegraphics[bb=54 350 350 600]{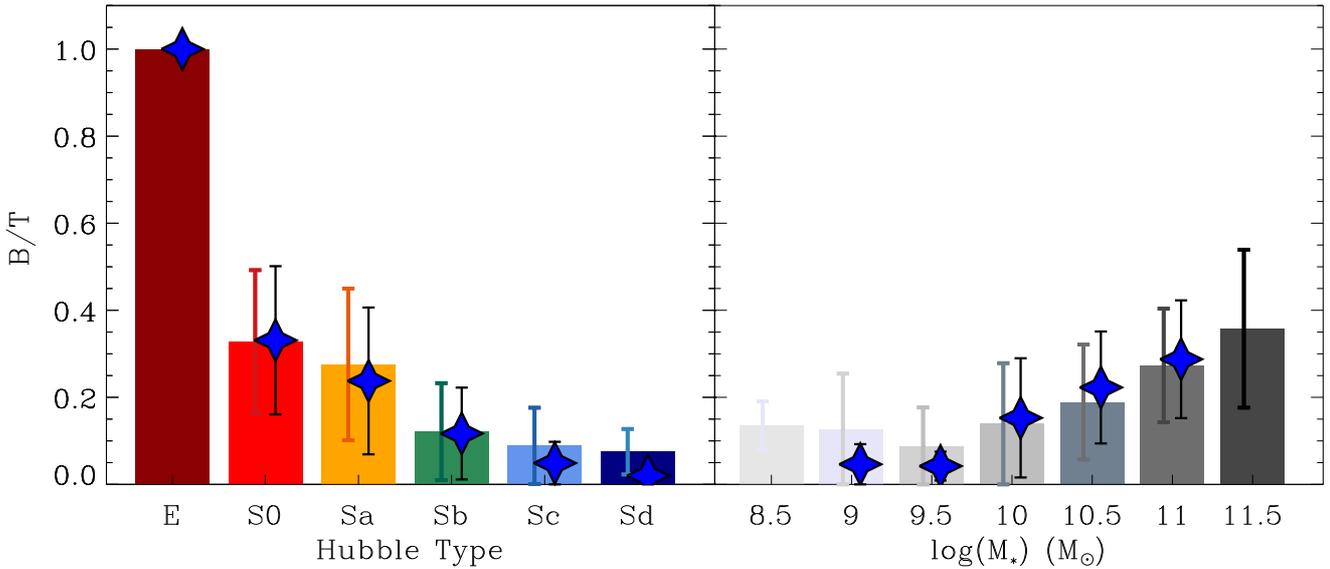}
\caption{Distribution of the $B/T$ ratio with Hubble type (left panel)
  and  stellar mass  (right panel).   Colour bars  represent the  mean
  values of  $B/T$ for the  observed distributions. Lines  with darker
  colours  show  the  1$\sigma$  error. Blue  stars  show  the  volume
  weighted  mean values  of the  $B/T$ ratios  with its  corresponding
  1$\sigma$ errors.   Elliptical galaxies have been  excluded from the
  distribution in the right panel. Bins  with less than five galaxies are
  not represented.}
\label{fig:bt}
\end{figure*}

The differences of this work with  respect to previous studies lie not
only  in the  different  techniques  (1D vs.   2D),  but  also in  the
different   samples   under   analysis.   The   sample   analysed   in
\citet{pohlentrujillo06}  consists  of  94 late-type  spiral  galaxies
(11\%  Type   I,  66  \%   Type  II,   and  33\%  Type   III)  whereas
\citet{erwin08} and  \citet{gutierrez11} analysed  66 barred  (27\% of
Type I, 42\% Type II, and 24\% Type III) and 47 unbarred (28\% of Type
I, 21\% Type II, and 51\% Type III) early-type galaxies, respectively.
Figure \ref{fig:breaks} shows that, according  to our 2D approach, the
fraction of Type I profiles  decreases with later Hubble types whereas
the fraction of  Type II profiles increases (with the  exception of Sd
galaxies)  in  agreement  with  previous findings  using  1D  analyses
(although with  different fractions).  For  Type III profiles  we find
that the fraction of galaxies displaying this profile remains constant
with Hubble  type.  No significant  trends are  found in terms  of the
stellar   mass  or   the  presence   of  a   bar  in   agreement  with
\citet{marino16}.   A   summary   of   our   results   is   shown   in
Table~\ref{tab:discfraction}.

\subsubsection{NPS and pure disc galaxies}
\label{sec:NPS}

There is compelling evidence that low $B/T$ and pure disc galaxies are
common  in the  Local Universe,  especially in  low-mass or  late-type
galaxies  \citep{boker02,salo15}.   Recent  works using  large  galaxy
samples drawn from  the SDSS survey have found that  15\%–20\% of disc
galaxies   out   to    $z   \sim$   0.03   appear    as   pure   discs
\citep{kautsch06,barazza08,kautsch09}.     The   high    observational
incidence  of both  pure discs  and low  $B/T$ galaxies  in the  Local
Universe is challenging for  cosmological galaxy formation simulations
to   reproduce  \citep{abadi03,governato04,peeblesnuser10}   and  only
recently,  galaxies  with  realistic  $B/T$  distributions  have  been
obtained \citep[see][for a review]{brooks16}.

\begin{figure}[!ht]
\includegraphics[width=0.49\textwidth]{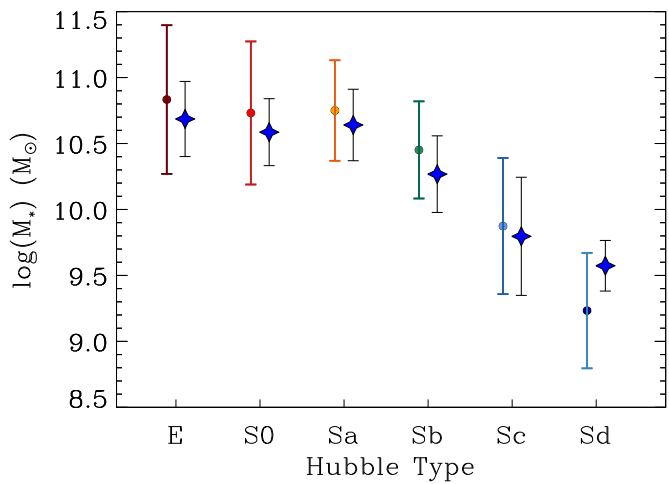}
\caption{Distribution of galaxy stellar mass  with Hubble type for our
  galaxy sample.  Colour points and bars represent the mean values and
  1$\sigma$ error of stellar mass for the observed distributions. Blue
  stars show  the volume-weighted mean  values of  the mass  with its
  corresponding 1$\sigma$ errors. }
\label{fig:htvsmass}
\end{figure}

Figure~\ref{fig:bt} shows the  distribution of $B/T$ as  a function of
the Hubble type and galaxy mass.   The relative size of the bulge with
respect  to   the  galaxy  is   one  of  the  primary   features  that
distinguishes different Hubble types, therefore the drop of $B/T$ from
early- to  late-type galaxies is  not surprising.  Our result  is also
quantitatively   in    agreement   with   previous   works    in   the
literature. \citet{laurikainen10}  found a  decline in the  mean $B/T$
values from 0.32 to 0.07 for S0 and Sd galaxies, respectively. Similar
values  were   more  recently  obtained  by   \citet{salo15}  and  are
consistent with our volume-corrected measurements of 0.33 and 0.02 for
S0 and Sd galaxies, respectively (see Table~\ref{tab:btfraction} for a
summary of the $B/T$ values). We also  find a clear trend of the $B/T$
ratio increasing with the galaxy stellar mass for the volume-corrected
sample  (right panel  on Fig.~\ref{fig:bt}).   Nonetheless, these  two
relations are  not completely  independent since, in our  sample, later
Hubble  types are  systematically less  massive than  early-types (see
Fig.~\ref{fig:htvsmass}).   This trend  between  the  Hubble type  and
stellar mass holds for the observed, but also for the volume corrected
quantities,  and   has  already  been  found   in  the  literature
\citep{huertascompany11,torrespapaqui12}.

\begin{table}[!t]
\caption{Volume corrected $B/T$ distributions in our galaxy sample}
\begin{center}
\begin{tabular}{l c}
\hline
Hubble Type  &  $\langle B/T \rangle$\\
\hline
S0   & 0.33$\pm$0.16 \\
Sa   & 0.24$\pm$0.17  \\
Sb   & 0.12$\pm$0.11  \\
Sc   & 0.05$\pm$0.09 \\
Sd   & 0.02$\pm$0.05 \\
\hline
Stellar Mass & $\langle B/T \rangle$\\
\hline
$9 < {\rm log(M_{\star}/M_{\sun})}>9.5$   & 0.05$\pm$0.13 \\
$9.5 < {\rm log(M_{\star}/M_{\sun})}>10$  & 0.05$\pm$0.09  \\
$10 < {\rm log(M_{\star}/M_{\sun})}>10.5$ & 0.15$\pm$0.14  \\
$10.5 < {\rm log(M_{\star}/M_{\sun})}>11$ & 0.22$\pm$0.13 \\
$11 < {\rm log(M_{\star}/M_{\sun})}>11.5$ & 0.29$\pm$0.13 \\
\hline
\end{tabular}
\end{center}
\label{tab:btfraction}
\end{table}

We find that 24\% and 76\%  of our volume-corrected disc galaxies have
$B/T>0.2$  and  $B/T<0.2$,  respectively.  The  relative  fraction  of
galaxies with different $B/T$ imposes strong constraints on the galaxy
evolution scenarios.   \citet{weinzirl09} found that 34\%  and 66\% of
their high-mass (${\rm log(M_{\star}/M_{\sun})}>10$) spiral sample had
$B/T>0.2$  and $B/T<0.2$,  respectively. Limiting  our sample  to this
mass limit we find 36\% and 64\% of our volume-corrected disc galaxies
have  $B/T>0.2$  and  $B/T<0.2$,   respectively,  in  remarkably  good
agreement with their results.

\begin{figure*}[!ht]
\includegraphics[width=\textwidth]{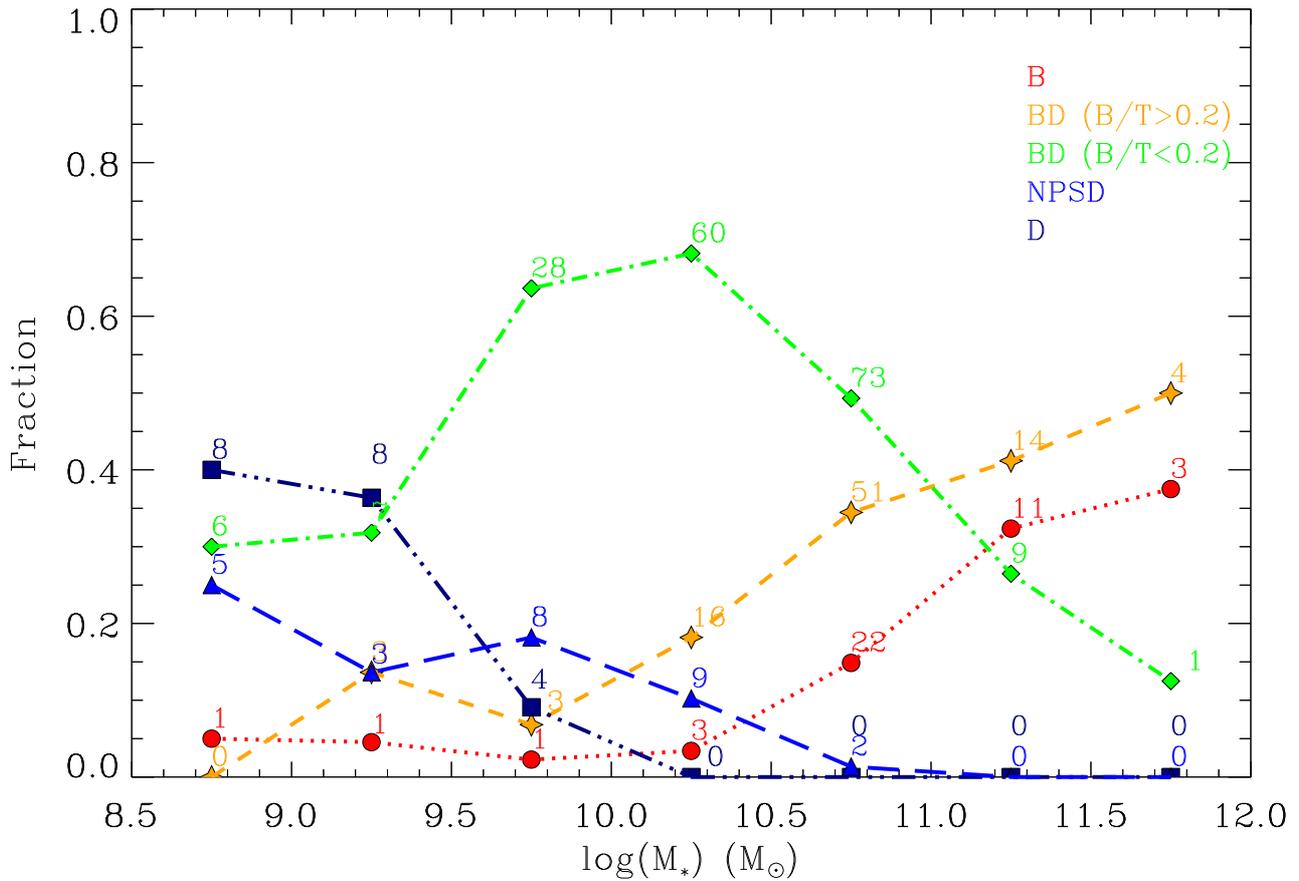}
\caption{Fraction  of  the  final   models  used  in  the  photometric
  decomposition as a function of  stellar mass.  Red circles represent
  single S\'ersic models (B).  Orange  stars show models composed of a
  bulge+disc  (BD)  with  $B/T>0.2$.  Green  diamonds  display  models
  composed of  a bulge+disc  with with  $B/T<0.2$.  Navy  blue squares
  show  models  with no  bulge  but  pure  disc (D).   Blue  triangles
  represent  models  with  a  NPS+disc  (NPSD).   The  five  different
  combination of  structures include broken profiles  and/or bars. The
  number of galaxies in each bin is also shown.}
\label{fig:models}
\end{figure*}

Figure~\ref{fig:models} shows the fraction  of the final decomposition
models  used to  represent  our galaxy  sample as  a  function of  the
stellar mass.  Here  we focus on the incidence of  bulge components in
the  photometric decomposition,  and therefore  we group  together all
possible  model  combinations  (see Table~\ref{tab:tabfits})  in  five
groups depending only on their central component, and neglecting other
structures such as broken discs  or bars: i) galaxies well represented
by a single  S\'ersic model (42 ellipticals),  ii) bulge+disc galaxies
with $B/T>0.2$ (91 galaxies),  iii) bulge+disc galaxies with $B/T<0.2$
(184  galaxies),  iv)  galaxies  with  no  bulge  and  pure  disc  (22
galaxies),    and     v)    galaxies     with    a     NPS+disc    (27
galaxies). Figure~\ref{fig:models} shows a clear segregation of galaxy
types depending on their stellar mass. The fraction of single S\'ersic
(i.e.  ellipticals) and  bulge+disc with  $B/T>0.2$ galaxies  steadily
increase with galaxy mass.  Indeed they are the dominant population of
galaxies  with $M_{\star}  > 10^{11}  M_{\sun}$.  On  the other  hand,
bulge+disc  with  $B/T<0.2$  galaxies   are  the  preponderant  galaxy
population in the mass range $10^{9.5} < M_{\star}/M_{\sun} <10^{11}$,
whilst galaxies  with negligible bulges,  both pure disc  and NPS+disc
models, prevail at low masses $M_{\star} < 10^{9} M_{\sun}$.

\section{Quality assessment and uncertainties in the photometric decomposition}
\label{sec:uncertainties}

Determining  the  uncertainties  in   the  photometric  parameters  in
multi-component decompositions such as the one presented in this paper
is a  complicated task. In this  Section, we describe our  attempts to
provide meaningful  uncertainties for  the different  parameters using
three  different  approaches: a  visual  quality  control analysis,  a
quantitative  error   measurement  based  on  mock   galaxies,  and  a
comparison with previous works in the literature.

\subsection{Visual quality check of the fits}
\label{sec:quality}

The galaxy sample analysed in this paper was drawn from the CALIFA DR3
sample by applying  two main criteria: i) they are  not interacting or
merging  and  ii)  they  are relatively  face-on  ($i  \lesssim  70$).
However, there  could still be  issues related to either  the original
SDSS imaging, the presence of  strong extra components not included in
the  fit (spiral  arms  or  rings), or  difficulties  inherent to  the
fitting  process  that  can  compromise   the  quality  of  the  final
decomposition. In order to assess the  reliability of the fits, and to
quantify  the incidence  of  these  issues in  our  galaxy sample,  we
carried out a visual check of each individual fit assigning a `quality
flag' that is provided in the  tables together with the results of the
decompositions.

The quality flag is defined as a number  (1, 2, 3, or 4) followed by a
letter ($a$, $b$,  or $c$).  Different numbers  correspond to galaxies
with (1) both good imaging and no other strong extra galaxy components
affecting the fit,  (2) good imaging but strong  extra components that
might influence  the final fit, (3)  poor imaging but no  strong extra
components affecting  the final fit,  and (4) poor imaging  and strong
extra components  affecting the final  fit. We refer to  problems with
the original  imaging as  those due  to the  presence of  other bright
galaxies whose stellar  haloes overlap with our  galaxy, and/or strong
fluctuations  of the  local  sky background  around  the galaxy.   The
different  letters are  related to  the fitting  process.  Due  to the
highly  degenerate parameter  space  we are  dealing  with (some  fits
include up to  17 free parameters), in some cases  it was not possible
to achieve a reasonable fit without  fixing some parameters to a value
given by our 1D analysis. We assigned the letter $a$ to fits where all
the parameters are  free to vary during the fitting  process.  If only
geometrical parameters such as PA or $\epsilon$ are kept fixed then we
assigned  a $b$,  and  whenever  we also  needed  to fix  size-related
quantities  we  considered   them  as  a  $c$.    This  quality  check
classification scheme  for each galaxy  was performed by at  least two
reviewers. In the  case of disagreement, another  reviewer checked the
quality flag  and provided  the final  classification.  The  number of
galaxies with each particular flag is shown in Table~\ref{tab:flag}.

\begin{table}
\caption{Results from the visual quality check of the photometric decompositions.}   
\label{tab:flag}    
\centering            
\begin{tabular}{c c c }      
\hline\hline              
Flag number   &  Flag letter  &  Number  \\
   (1)        &    (2)        &    (3)   \\
\hline
              &     $a$       &  190     \\
1             &     $b$       &  14      \\
              &     $c$       &  122     \\
\hline 
              &     $a$       &  22      \\
2             &     $b$       &  5       \\
              &     $c$       &  26      \\
\hline 
              &     $a$       &   13     \\
3             &     $b$       &   0      \\
              &     $c$       &   10     \\
\hline 
              &     $a$       &   1      \\
4             &     $b$       &   0      \\
              &     $c$       &   1      \\
\hline                                 
\end{tabular}
\tablefoot{(1)  Different numbers  represent galaxies  with both  good
  imaging and no other strong  extra components (spiral arms or rings)
  affecting the fit (flag 1), good imaging but strong extra components
  that  can influence  the final  fit (flag  2), poor  imaging but  no
  strong extra  components affecting the  final fit (flag 3)  and poor
  imaging and strong extra components  (flag 4); (2) Different letters
  represent whether  or not: ($a$)  all the parameters are  allowed to
  vary during  the fitting process, ($b$)  only geometrical parameters
  such  as PA  or $\epsilon$  are kept  fixed, and  ($c$) size-related
  quantities are also fixed.}
\end{table}

Only  $\sim$6\% of  the images  were  classified as  poor quality  and
$\sim$ 14\%  present strong extra components  that might significantly
affect  the fit.   On  the other  hand, $\sim$58\%  of  the fits  were
performed allowing  all the parameters  to vary during the  fit.  This
highlights the high level of  convergency in our fits since $\sim$73\%
of our  models require the  constraint of $>10$  parameters. Regarding
those fits  with fixed quantities, only  5\% of the fits  required the
ellipticities or PA of any of the components to be fixed to the values
obtained from the 1D radial profiles.  However, $\sim$37\% of the fits
have been  performed fixing a  size-related quantity of the  model. We
find that in approximately half of these cases the fixed parameter was
the break radius of the disc in the Type II and Type III profiles.  As
explained  in Sect.\ref{sec:breaks},  constraining  the parameters  of
disc breaks in a multi-component decomposition is not straightforward,
but it is  the only way to consistently compare  with the other galaxy
structures. It is  worth noticing that the fixing of  the break radius
does  not  affect   our  classification  nor  the   results  shown  in
Sect.\ref{sec:breaks}.

\subsection{Statistical errors based on mock galaxy simulations}
\label{sec:mock}

The formal errors obtained from the minimisation procedure are usually
not  representative  of  the  real errors  on  the  fitted  parameters
\citep{mendezabreu08}.   This is  mainly because  possible covariance
terms are neglected  in the error computation  process.  Therefore, to
provide  our structural  parameters with  a robust  error estimate  we
carry  out different  tests applying  Monte Carlo  techniques to  mock
galaxies. This procedure allows us to improve both the accuracy of our
error estimates,  avoiding the otherwise necessary  exploration of the
full parameter space to account for the parameters' covariance, and to
determine the observational limits of our photometric decomposition.

We devised  a set of  tailor-made simulations for each  combination of
structures     fitted     to      our     sample     galaxies     (see
Table~\ref{tab:tabcomponents}).  For  each set,  we simulate  500 mock
galaxies with  structural parameters constrained within  the limits of
our real  galaxy sample.  For  the sake  of simplicity we  assumed the
$i-$band parameters as representative  of our galaxies.  Altogether we
have  created 7000  mock  galaxies  that are  built  in the  following
way. First,  the total  apparent magnitude of  the galaxy  is randomly
selected within the observed range  $m_i$ = [11,14].  Then, the values
of the relative luminosity ratios  of the bulge ($B/T$), disc ($D/T$),
bar ($Bar/T$),  and NPS  ($NPS/T$) are set  according to  the observed
values for  each combination of  structures.  The distribution  of the
luminosity ratios also matches that of  the real galaxies, and a given
luminosity is associated  to each structure.  The  interval ranges for
the  size and  shape  parameters  for each  structure  covered by  our
simulations  are:  $r_{\rm  e}$ =  [0.9$\arcsec$,14$\arcsec$],  $n$  =
[0.5,5],   $h$   =   [4$\arcsec$,25$\arcsec$],   $r_{\rm   break}$   =
[16$\arcsec$,50$\arcsec$],  $h_{\rm out}$  = [4$\arcsec$,25$\arcsec$],
$r_{\rm  bar}$  =  [5$\arcsec$,30$\arcsec$].   Once  the  mock  galaxy
structure is  defined, the  values of the  surface brightness  for the
bulge ($\mu_{\rm e}$), disc  ($\mu_{\rm 0}$), bar ($\mu_{\rm 0,bar}$),
and NPS ($\mu_{\rm 0,NPS}$) are derived using the equations to compute
the total luminosity for each  analytical representation of the galaxy
structure. Finally,  the axis  ratio of  the bulge  ($q_{\rm bulge}$),
disc ($q_{\rm  disc}$) and bar  ($q_{\rm bar}$) components as  well as
the value of the position angle  of the bulge (PA$_{\rm bulge}$), disc
(PA$_{\rm disc}$) and bar (PA$_{\rm bar}$) are generated from a random
uniform  distribution,  where  no  constraints are  adopted.   In  the
particular case of  the double barred galaxies the  typical values for
the mock galaxy generation were  obtained from de Lorenzo-C\'aceres et
al.  ({\it in prep}).
  
Mock galaxies are  placed at a distance of 67  Mpc that corresponds to
the  median value  of  our real  sample. The  galaxy  models are  then
convolved  with  the  mean  PSF  of  the  $i-$band  SDSS  images  (see
Sect.~\ref{sec:psf}) to reproduce the  typical spatial resolution.  In
addition, we adopt the pixel  scale (0.396 arcsec/px), and the typical
values  of the  CCD gain  (4.86  e$^-$/ADU) and  read-out noise  (5.76
e$^-$) to mimic  the instrumental setup of the SDSS  data.  Finally, a
background  and photon  noise are  added to  the artificial  images to
yield a signal-to-noise ratio to match the observed one.

The mock  galaxies are then fitted  using GASP2D as if  they were real
galaxies. The  comparison between the  input and output values  of the
fitted parameters is  used to compute the errors.   Mock galaxies were
split   into  three   different   bins   of  magnitude   ($11<m_i<12$,
$12<m_i<13$,  $13<m_i<14$). In  each bin,  the mean  and the  standard
deviation of  the relative  errors are adopted  as the  systematic and
statistical  errors  for  the  observed galaxies  for  the  parameters
($\mu_{\rm e}$, $r_{\rm  e}$, $n$, $\mu_{\rm 0}$,  $h$, $h_{\rm out}$,
$r_{\rm break}$, $\mu_{\rm 0,bar}$, $a_{\rm bar}$, $\mu_{\rm 0,NPS}$),
whereas the  mean and  standard deviation of  the absolute  errors are
adopted  as the  systematic and  statistical errors  for the  observed
galaxies  for  the set  ($q_{\rm  bulge}$,  PA$_{\rm bulge}$,  $q_{\rm
  disc}$, PA$_{\rm  disc}$, $q_{\rm bar}$, PA$_{\rm  bar}$). Since the
systematic errors  are generally  smaller than the  statistical errors
they were added in quadrature. Errors are then assigned to each galaxy
depending  on  its apparent  magnitude.   The  final errors  for  each
parameter are included in Table~\ref{tab:tabfits}.

\subsubsection{Limitations to the measurement of breaks in disc galaxies}
\label{sec:breaklimits}

\begin{figure}[!ht]
\includegraphics[width=0.49\textwidth]{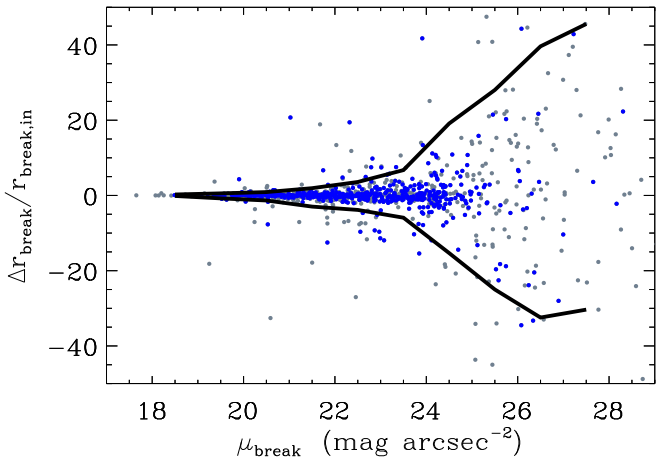}
\caption{Relative  percentage error  in the  measurement of  the break
  radius ($(r_{\rm  break,in} - r_{\rm  break,out})/r_{\rm break,in}$)
  as a  function of  the surface brightness  at $r_{\rm  break}$.  The
  individual points represent all the mock galaxies with a broken disc
  independently of the presence of  other components. Type II and Type
  III  disc   profiles  are   shown  with   grey  and   blue  circles,
  respectively.  The thick lines enclose 90\% of the galaxies.}
\label{fig:breaklimits}
\end{figure}

The mock  galaxy simulations  described in  the previous  section also
allow us to quantify the limitations  for a robust measurement of disc
breaks. To this end, we combined the six sets of simulations involving
the presence of  a broken disc independently of the  presence of other
components  (see  Table~\ref{tab:tabcomponents}).   The  final  sample
includes  3000 mock  galaxies.   Fig.~\ref{fig:breaklimits} shows  the
relative error in the measurement of the break radius as a function of
the surface brightness where the break occurs. It is worth noting that
both Type II and Type III  profiles are represented in the simulations
and they are  shown with grey and blue circles,  respectively. We find
that  disc  breaks  can  be  robustly  measured  up  to  a  SB  of  24
mag/arcsec$^2$  (within a  $\sigma \sim$  3\%).  However,  beyond this
limit the determination of the break radius is more uncertain becoming
completely undefined  for $\mu_{\rm break}>25$  mag/arcsec$^2$.  These
simulations demonstrate the  limits of the SDSS  imaging in accurately
measuring breaks in disc galaxies.

\subsection{Comparison with previous CALIFA data}
\label{sec:compCALIFA}

A simple  approach to understanding  the structure of galaxies  is the
growth curve  analysis. \citet{walcher14}  analysed the  CALIFA mother
sample  using  this  methodology  on  SDSS-DR7  images.  They  derived
reliable values  of the total  magnitude, half-light major  axis, axis
ratios, and position angles of the CALIFA galaxies. These measurements
have been  used in several  works within the CALIFA  collaboration and
therefore it is  instructive to compare them with the  results of this
paper.

\begin{figure}
\includegraphics[bb=20 340 200 800]{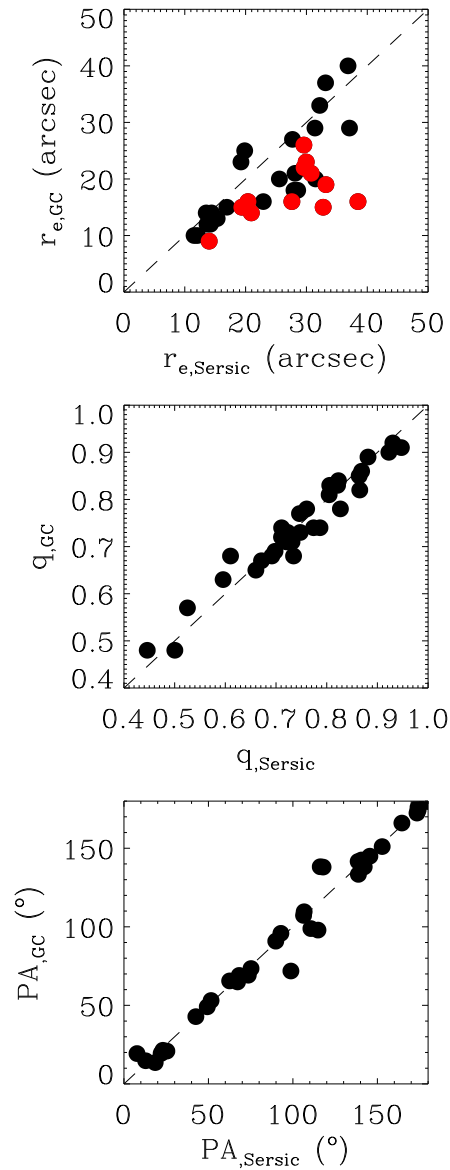}
\caption{Comparison  of  our photometric  decomposition  \emph{r}-band
  estimated parameters (labelled as S\'ersic) with those obtained from
  the growth curves (labelled as GC) in \citet{walcher14}. Only those
  galaxies fitted  with a  single S\'ersic profile  have been  used in
  this  comparison.   From  top  to  bottom,  panels  show  the  galaxy
  effective radius, axis ratio, and position angle, respectively.  Red
  points in  the upper  panel represent galaxies  fitted with  a single
  S\'ersic with $n>5.5$.}
\label{fig:compCALIFA}
\end{figure}

Since  growth curve  analysis does  not separate  different structural
components  but considers  the galaxy  a single  entity, a  meaningful
comparison can only be achieved  for our galaxies classified as single
S\'ersic models (B). Fig.~\ref{fig:compCALIFA} shows the values of the
effective radius  ($r_e$), axis  ratio ($q$)  and position  angle (PA)
obtained  through  the  growth  curve  and  photometric  decomposition
analysis. Different methods show very good agreement in terms of the
geometrical parameters, with no  systematic variations and an absolute
scatter of  0.03 and 4  degrees in $q$  and PA, respectively.   On the
contrary,  the values  of  $r_e$  do show  a  systematic offset,  with
shorter values measured using the  growth curve analysis.  This result
is  expected since  the S\'ersic  parameterisation assumes  the galaxy
model is extended to infinity, and  therefore part of the galaxy light
is still present in the outer  and low surface brightness wings of the
profile.  In Fig.~\ref{fig:compCALIFA} we have labelled those galaxies
with  a  S\'ersic index  $n>5.5$  in  red.   We  show how  the  larger
deviations actually appear  in these galaxies where  the profile wings
are more  prominent. The relative  scatter between the two  methods is
$\sim$20\%.

\subsection{Comparison with the literature}
\label{sec:literature}

\begin{figure}
\includegraphics[]{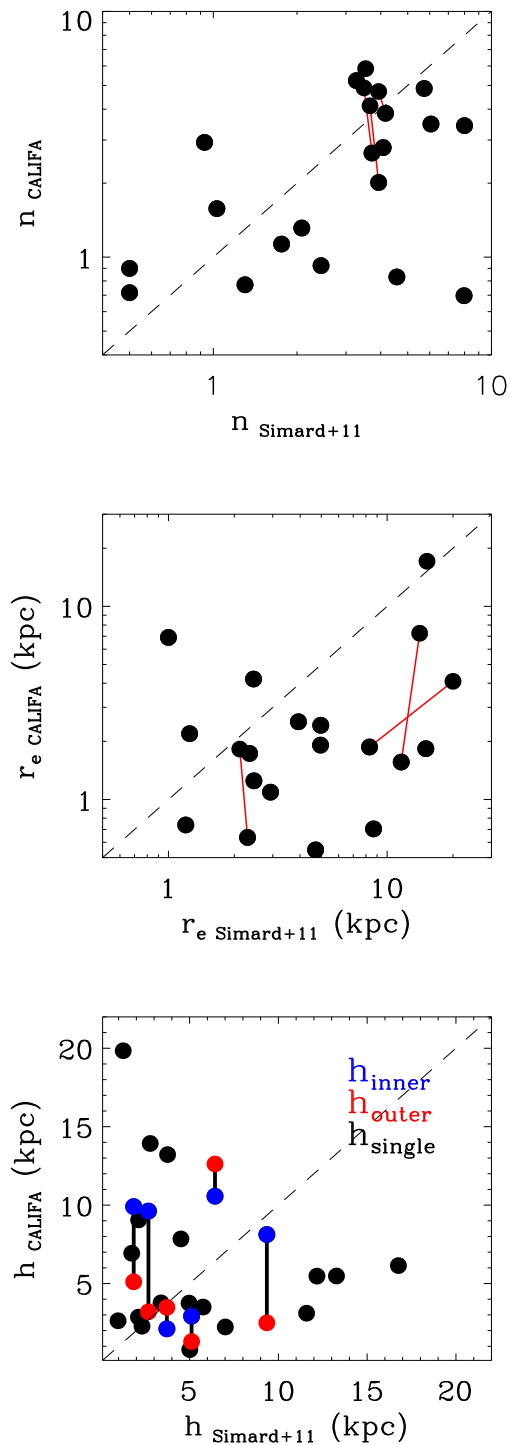}
\caption{Comparison  of our  \emph{r}-band  estimated parameters  with
  those    from   the    GIM2D    photometric   decompositions    from
  \citet{simard11}.  Top  and middle panels show  bulge S\'ersic index
  and  effective radius  respectively.  Parameters  from the  one- and
  two-component fits for the  early-type galaxies with B/BD morphology
  are linked with solid red lines.  The bottom panel compares the disc
  scale lengths.  Inner and outer disc scale lengths are distinguished
  in the case of breaks as blue and red dots, respectively, and linked
  with a solid  black line. \citet{simard11} do not  take into account
  breaks in their analysis.}
\label{fig:plotsimard}
\end{figure}

\begin{figure*}[!ht]
\includegraphics[angle=90,width=\textwidth]{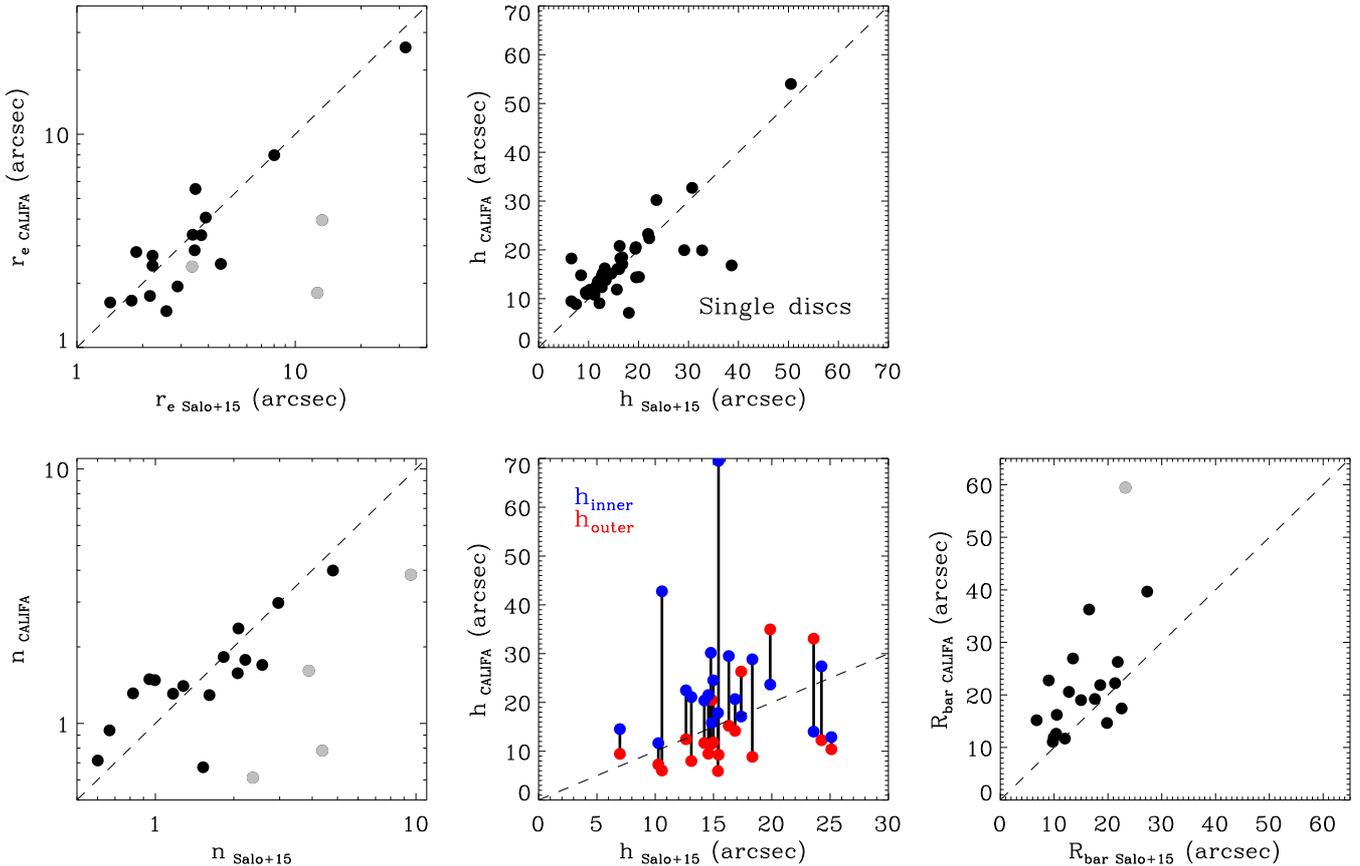}
\caption{Comparison  of  our  \emph{r}-band best-fit  parameters  with
  those  from the  multi-component photometric  decompositions of  the
  3.6\,$\mu$m images of the S$^4$G galaxies \citep{salo15}.  There are
  65 galaxies  in common between  our sample and S$^4$G.   Left panels
  correspond  to elliptical/bulge  parameters: effective  radius (top)
  and  S\'ersic index  (bottom). Controversial  bulges (see  text) are
  plotted as grey dots.  Middle  panels compare the disc scale lengths
  for single  discs (top)  and truncated discs  as classified  by this
  work (bottom).   \citet{salo15} only  include single discs  in their
  decompositions.    Inner   and   outer  disc   scale   lengths   are
  distinguished  in  the  case  of   breaks  as  blue  and  red  dots,
  respectively. For  the sake of  clarity, the 1:1 axis  proportion in
  the  bottom panel  has not  been kept.   The right  panel shows  the
  comparison  of the  bar lengths  for the  19 galaxies  classified as
  barred by both works.  The use  of two different disc components (an
  exponential and  Ferrers) in the  S$^4$G fit explain  the difference
  for NGC\,5957 (grey dot).}
\label{fig:plots4g}
\end{figure*}

Given the  human-supervised nature  of the  photometric decompositions
presented here, the choice of the number of components included in the
fit is a relevant source of  uncertainty which is difficult to account
for.  Two-component (bulge+disc) models are the usual approach to deal
with large datasets in an automatic way. In these cases the error bars
associated with certain  parameters might be excessively  large due to
the    absence     of    critical    structures    such     as    bars
\citep[e.g.][]{gadotti08}.    Best    suited   multi-component   fits
alleviate this problem but introduce the human factor. In this section
we  present a  comparison  of  our results  against  two recent  works
comprising large  datasets: \citet{simard11},  who apply  an automatic
up-to-two-component   fitting   procedure   to   the   SDSS-DR7,   and
\citet{salo15}, who  perform human-supervised multi-component  fits of
the whole S$^4$G  sample \citep{sheth10}.  The main  results are shown
in Figs.~\ref{fig:plotsimard} and \ref{fig:plots4g}, respectively.

\citet{simard11}  analyse the  \emph{g}- and  \emph{r}-band images  of
over a  million galaxies retrieved  from the SDSS-DR7. They  apply the
GIM2D  code   \citep{simard98}  to  perform  automatic   one-  (single
S\'ersic) and  two-component (bulge+disc) decompositions of  the whole
sample, by  using the information  from the two  bands simultaneously.
In  Fig.~\ref{fig:plotsimard} we  plot some  relevant bulge/elliptical
and  disc parameters  as obtained  from  both their  approach and  our
multi-component  fitting procedure  in \emph{r}-band.   The number  of
components is consistently taken into account, so the galaxies that we
classify   as  ellipticals   (see  Section~\ref{sec:onevsmulti})   are
compared with the one-component fit from \citet{simard11}.  Otherwise,
the bulge+disc decomposition is  considered.  Both options are plotted
for      the      cases      with     B/BD      morphologies      (see
Section~\ref{sec:onevsmulti}).     It    is    worth    noting    that
\citet{simard11} also provide two-component fits in which the S\'ersic
index is fixed to $n=4$.  Those fits are not used in this comparison.

There   are   25  galaxies   in   common   between  our   sample   and
\citet{simard11}.   Galactic  location on  the  sky  explains the  low
number  of common  galaxies, as  many  CALIFA galaxies  belong to  sky
regions   not   included  in   the   SDSS   Legacy  Survey   used   by
\citet{simard11}.  The large dispersion shown in the plots is expected
as \citet{simard11} do  not include other relevant  structures such as
bars,  inner  discs, or  main  disc  truncations  in their  fits.   In
general,   the  bulge   parameters   ($n$  and   $r_e$)  measured   by
\citet{simard11} tend to be larger than those from this work, as their
bulges need to account for other central structures as well.

The S$^4$G sample consists of  2352 galaxies observed with the Spitzer
Infrared   Array  Camera   (IRAC)   in  the   near-infrared  3.6   and
4.5\,{$\mu$}m    bands.     GALFIT   \citep{peng02}    multi-component
photometric  decompositions  for  all  the  3.6\,{$\mu$}m  images  are
presented in \citet{salo15}.  The  strategy followed in their analysis
is   analogous  to   the  one   shown  in   this  paper.    There  are
notwithstanding  some important  differences that  are worth  pointing
out.   First,  \citet{salo15} do  not  use  truncated discs  in  their
models; instead  only single exponential  discs are fitted.   If there
are clear differences  between the inner and outer  disc regions, they
include an  extra component that  it is added  up to the  others. This
component is tagged as a disc but can have a Ferrers or Sérsic profile
if that better  resembles the image. S$^4$G galaxies  can therefore be
classified  as  having  multiple  disc structures  when  only  one  is
actually present.   In other  cases however, the  galaxy can  have two
actual discs (e.g.  the usual disc plus an inner disc).  Secondly, for
very  inclined galaxies,  \citet{salo15} fit  three-dimensional discs.
Whereas the usual exponential disc is considered infinitesimally thin,
they include {`edgediscs'} with  vertical extension for those galaxies
with $q  \lesssim 0.2$.   Since we apply  a threshold  in inclination,
none of  the CALIFA galaxies  in common  with S$^4$G have  been fitted
with an {edgedisc}.
 
The  human-supervised choices  of components  from \citet{salo15}  and
this work match exactly  for 47 out of the 65  galaxies common to both
samples.   For the  reasons stated  above, we  consider multiple  disc
structures in  the S$^4$G  classifications as  equivalent to  one disc
component in the CALIFA galaxies.  Similarly, NPS and bulge components
are    equally    taken    into     account.     As    indicated    in
Section~\ref{sec:function}, and also discussed in \citet{salo15}, NPSs
are  usually unresolved  bulges.  For  these cases,  it is  always the
S$^4$G classification  that includes  a NPS, while  a proper  bulge is
shown in the CALIFA analysis, as  expected due to the lower resolution
of the near-infrared images when compared to SDSS data.

The fact that the classifications from both teams, which use different
codes  and  images,  match  for  72\% of  the  common  sample,  is  an
indication of  the robustness of the  human-supervised multi-component
photometric  decomposition techniques.   The 19  non-matching galaxies
show relevant differences in the classifications, such as the presence
of  bulges or  bars. In  particular, we  include bars  in five  CALIFA
galaxies which are unbarred for  the S$^4$G team, whereas the opposite
occurs for seven galaxies.  Since it  has been found that the measured
bar     fraction    increases     in    near-infrared     observations
\citep[e.g.][]{marinovajogee07}, a higher bar occurrence in the S$^4$G
data is expected.  No significant  differences are found in the common
sample of 65 galaxies.  Regarding the early-types, there is one CALIFA
elliptical galaxy classified as S0 by S$^4$G and viceversa.

Figure~\ref{fig:plots4g}  shows the  comparison  of  bulge (or  single
S\'ersic), disc,  and bar parameters between  our \emph{r}-band images
and  the 3.6\,{$\mu$}m  images used  by \citet{salo15}.   Overall, the
results are in good agreement.  We have visually explored the outliers
shown  in  the  plots  for  the bulge  parameters  (Sérsic  index  and
effective radius).   Some correspond  to bulges  that acquire  high SB
values in the  outer galaxy regions, even surpassing  the disc profile
in the  \citet{salo15} fits.   This is  considered unrealistic  in our
decomposition  and  therefore they  show  too  high Sérsic  index  and
effective radius  values with respect  to the results from  this work.
In other cases,  different bar fits or  even the absence of  a bar for
one of the galaxies accounts for inconsistencies in bulge parameters.

Twenty  galaxies  of  the  common  sample  host  truncated  discs,  as
classified  in this  work. \citet{salo15}  do not  include breaks  and
truncations  in their  decomposition.  Truncations  can be  related to
structures such as rings or bars and it is therefore not obvious which
disc region, inner or outer, is  best suited for the comparison with a
single  disc  such  as  the  one provided  for  the  S$^4$G  galaxies.
Moreover,  the estimated  properties  of  a single  disc  may also  be
affected  by other  structures or  even image  parameters such  as sky
background,  thus  are  biased  towards inner  or  outer  measurements
depending on  each case. \citet{salo15}  show that their  single discs
are a good  estimate of a `mean'  disc of the inner  and outer values.
They also  find that outer discs  in Type II galaxies  better resemble
the properties  of actual single discs.   In Fig.~\ref{fig:plots4g} we
compare both  the inner and outer  disc scale lengths derived  in this
work with the single S$^4$G scale  lengths. Values from the outer disc
regions lie slightly closer to  the unity line, although the situation
is different  for each  galaxy.  Results  for galaxies  hosting single
discs as  concluded by both  CALIFA and S$^4$G classifications  are in
full agreement.

There are 19 galaxies classified  as barred by both \citet{salo15} and
this work  (bottom right  panel of Fig.~\ref{fig:plots4g}).   The main
outlier corresponds  to NGC\,5957.   \citet{salo15} fit a  bulge, bar,
and two discs  to this galaxy. While the main  disc has an exponential
profile, the second one is fitted with a Ferrers profile. In this work
NGC\,5957  is fitted  with  a bulge,  bar, and  single  disc. We  have
verified that our bar corresponds to  their Ferrers disc, and thus the
inconsistency in the  results, while we do not find  signatures of any
other inner bar in the optical images.\\

\section{Conclusions}
\label{sec:conclusions}

This paper  presents the multi-component photometric  decomposition of
404 galaxies  from the CALIFA  DR3 survey. Our  aim is to  provide the
community  with  an  accurate   photometric  characterisation  of  the
multiple  stellar  structures  shaping  the  CALIFA  galaxies;  namely
bulges, bars, and discs.

The galaxy  sample covers  all galaxies included  in the  CALIFA final
data release,  both from the  mother sample and the  extension sample,
except  those in  mergers,  interactions, and  those  that are  highly
inclined. We  adopted a  human-supervised strategy  where up  to three
structural   components:  bulge/NPS   (nuclear  point   source),  disc
(including breaks), and bar are used  to provide the best fit for each
galaxy.   The   final  combination  of  structures   was  individually
determined by the code-user  after checking several possibilities.  We
have  thoroughly compared  our results  with previous  works from  the
literature  \citep{simard11,salo15} obtaining  a reasonable  agreement
when  the  different galaxy  components  are  carefully selected,  but
different results when  automatic methods are used.   We consider this
as   an  indication   of  the   robustness  of   the  human-supervised
multi-component photometric  decomposition techniques.   Counting disc
breaks as different  structures, we used 13  different combinations of
structures to describe our sample galaxies.

We  focused  on the  incidence  of  the  different structures  in  our
sample.  Since  the galaxies  extracted  from  the mother  sample  are
volume-correctable,  we have  studied the  frequency of  the different
structures for the  whole observed sample and  volume -corrected.  Our
main conclusions are:

\begin{itemize}
\item We found an average bar  fraction in our volume-corrected sample
  of  57\%,  which   is  consistent  with  previous   results  in  the
  literature.  The volume-corrected bar fraction shows  a drop toward
  late type  galaxies; unfortunately,  the number  of galaxies  in the
  late-type  bin is too  small to draw statistical  conclusions. The
  observed  bar  fraction  (using  the  whole  sample)  is  relatively
  constant with Hubble type.  Regarding the mass dependance of the bar
  fraction, the volume-corrected bar fraction drops from $\sim$75\% at
  $M_{\star}     =10^{9.5}$     $M_{\sun}$    to     $\sim$25\%     at
  $M_{\star}=10^{11}$ ${\rm M_{\sun}}$.

\item We  explored the frequency of  different disc types by  using 2D
  surface-brightness models including  broken exponential profiles and
  found that 62\%, 28\%, and 10\%  of our volume corrected disc sample
  are better represented  with a Type I (pure exponential),  a Type II
  (down-bending), and  a Type III (up-bending)  profile, respectively.
  These fractions are in strong  disagreement with previous results in
  the  literature  \citep{erwin08,gutierrez11}.   We  argue  that  the
  different  methodologies   are  the   main  explanation   for  these
  differences.  In our  2D analysis we are  simultaneously fitting all
  different  galaxy components  whereas  1D studies  usually fit  only
  piecewise  exponentials to  pre-defined regions  of the  SB profile.
  Despite  the  quantitative differences,  we  found  the same  trends
  observed in the previous works, that  is, a decrease in the fraction
  of Type I profiles with Hubble type (from Sa to Sc), an increase for
  the Type  II profiles and an  almost constant fraction for  type III
  galaxies. No  significant trends are  found in terms of  the stellar
  mass or the presence of bars.

\item  We  also studied  the  incidence  of  pure discs  and/or  small
  unresolved bulges in our sample.   Regarding the presence of a bulge
  and its  prominence, we find  a clear segregation of  the structural
  composition of  galaxies according to  stellar mass. At  high masses
  (${\rm  log(M_{\star}/M_{\sun})}>11$),  galaxies   modelled  with  a
  single  S\'ersic  or  a  bulge+disc  with  $B/T>0.2$  represent  the
  dominant    population.    At    intermediate   masses    ($9.5<{\rm
    log(M_{\star}/M_{\sun})}<11$), galaxies  described with bulge+disc
  but $B/T < 0.2$ are preponderant  whereas in the low mass end (${\rm
    log(M_{\star}/M_{\sun})}<9.5$),   the  prevailing   population  is
  constituted by galaxies  modelled with either pure  discs or nuclear
  point sources+discs (i.e.  no discernible bulge). This  trend of the
  fitted model with  galaxy mass is also consistent with  the trend of
  the $B/T$  ratio with both Hubble  type and mass.  We  found a clear
  decrease of  $B/T$ with both  increasing Hubble type  and decreasing
  mass with an average volume corrected $B/T$ value of 0.14.

\end{itemize}

The work presented in this paper focuses on describing the photometric
decomposition pipeline  and the incidence of  the different structural
components in  the final CALIFA  data release. It also,  however, sets
the basis for  new studies combining photometric  information with the
wealth of 2D spatially  resolved spectroscopic information provided by
the CALIFA survey.

\begin{acknowledgements}
JMA and  VW acknowledges  support from  the European  Research Council
Starting Grant (SEDmorph; P.I.  V. Wild). L.S.M acknowledges financial
support   from   the  Spanish   {\em   Ministerio   de  Econom\'ia   y
  Competitividad (MINECO)} via grant AYA2012-31935.  AdLC acknowledges
support from the  UK Science and Technology  Facilities Council (STFC)
grant  ST/J001651/1  and from  the  Spanish  Ministry of  Economy  and
Competitiveness (MINECO)  grant AYA2011-24728.  C.  C.-T.   thanks the
support  of the  Spanish  {\it Ministerio  de  Educaci\'on, Cultura  y
  Deporte} by means of the FPU fellowship program and the support from
the  {\it  Plan  Nacional  de Investigaci\'on  y  Desarrollo}  funding
programs, AYA2012-30717 and AyA2013-46724P, of Spanish {\it Ministerio
  de Econom\'ia y Competitividad}  (MINECO)'.  EF acknowledges support
by the MINECO grant AYA2014-53506-P and the Junta de Andaluc\'ia grant
FQM108.   JLA acknowledges  financial  support from  the MINECO  grant
AYA2013-43188-P. EMC and LC are  supported by Padua University through
grants   60A02-5857/13,   60A02-5833/14,    and   60A02-4434/15.    LC
acknowledges the University of St.   Andrews for the hospitality while
this paper was  in progress.  L.G.  was supported in  part by FONDECYT
through grant  3140566 and  the US  National Science  Foundation under
Grant AST-1311862.  IM acknowledges  financial support from the MINECO
grant  AYA-42227-P.  RAOM  acknowledges  support  from CAPES  (Brazil)
through a  PDJ fellowship  from project  88881.030413/2013-01, program
CSF-PVE.  PP is supported by FCT through the Investigador FCT Contract
No.   IF/01220/2013 and  POPH/FSE (EC)  by FEDER  funding through  the
program  COMPETE.    PSB  acknowledge  financial  support   from  the:
CONICYT-Chile Basal-CATA  PFB-06/2007 and the  AYA2013-48226-C3-1-P by
the Ministerio de  Ciencia e Innovacion.  This paper is  based on data
from  the  Calar  Alto  Legacy  Integral  Field  Area  Survey,  CALIFA
(http://califa.caha.es), funded  by the  Spanish Ministery  of Science
under    grant   ICTS-2009-10,    and    the   Centro    Astron\'omico
Hispano-Alem\'an.

Based  on observations  collected  at the  Centro Astronómico  Hispano
Alemán  (CAHA)  at Calar  Alto,  operated  jointly by  the  Max-Planck
Institut für Astronomie and the  Instituto de Astrofísica de Andalucía
(CSIC).     We    acknowledge     the    use     of    SDSS     data
(\url{http://www.sdss.org/collaboration/citing-sdss/}).
\end{acknowledgements}

\bibliographystyle{aa} 
\bibliography{reference} 

\end{document}